\documentclass[floats,floatfix,showpacs,amssymb,prd,twocolumn,superscriptaddress,nofootinbib,reprint,preprintnumbers]{revtex4-2}

\usepackage{amssymb,amsmath,verbatim,mathtools,needspace,enumitem,etoolbox,graphicx,physics,microtype,afterpage,xspace,tabularx,lmodern,multirow}
\usepackage{booktabs}
\usepackage{siunitx}
\usepackage{gensymb}
\usepackage{ulem}  
\usepackage[dvipsnames, usenames]{xcolor}
\definecolor{linkcolor}{rgb}{0.0,0.3,0.5}
\usepackage[unicode, colorlinks=true, linkcolor=linkcolor, citecolor=linkcolor, filecolor=linkcolor, urlcolor=linkcolor, linktocpage, breaklinks]{hyperref}
\usepackage[all]{hypcap}
\usepackage[T1]{fontenc}
\usepackage[utf8]{inputenc}
\usepackage[usenames,dvipsnames]{xcolor}
\hypersetup{colorlinks=true,citecolor=romared,linkcolor=romared,urlcolor=romared}

\definecolor{romared}{RGB}{142,0,28}

\newcommand{\be}{\begin{equation}}
\newcommand{\ee}{\end{equation}}

\def\be{\begin{equation}}
\def\ee{\end{equation}}
\newcommand{\beq}{\begin{eqnarray}}
\newcommand{\eeq}{\end{eqnarray}}

\definecolor{goldenrod}{RGB}{218, 165, 32}
\definecolor{navy}{RGB}{0, 0, 128}
\definecolor{tabblue}{RGB}{31, 119, 180}
\definecolor{tabcyan}{RGB}{0, 200, 210}
\definecolor{tabgray}{RGB}{100, 100, 100}
\definecolor{tabgreen}{RGB}{44, 160, 44}
\definecolor{tabred}{RGB}{214, 39, 40}
\definecolor{tabpink}{RGB}{255, 141, 161}
\definecolor{magenta}{RGB}{255, 0, 255}
\definecolor{darkmagenta}{RGB}{139, 0, 139}

\definecolor{tabpurple}{RGB}{148, 103, 189}
\usepackage{aas_macros}
\usepackage{makecell}
\usepackage{soul}

\usepackage{lipsum}
\usepackage{orcidlink}

\newcolumntype{Y}{>{\centering\arraybackslash}X}

\begin{document}
\preprint{\hbox{UTWI-22-2025}}
\title{Tracing the Neutrino-Induced Phase Shift in the 21-cm Spectrum}

\author{Gabriele Montefalcone\,\orcidlink{0000-0002-6794-9064}}\email{montefalcone@utexas.edu}
\affiliation{Texas Center for Cosmology and Astroparticle Physics,
Weinberg Institute for Theoretical Physics, Department of Physics,
University of Texas at Austin, Austin, TX 78712, USA}

\author{Hector Afonso G. Cruz\,\orcidlink{0000-0002-1775-3602}}
\affiliation{William H. Miller III Department of Physics and Astronomy, Johns Hopkins University, 3400 N. Charles Street, Baltimore, Maryland, 21218, USA}

\author{Julian B. Mu\~noz\,\orcidlink{0000-0002-8984-0465}} 
\affiliation{Department of Astronomy, The University of Texas at Austin, 2515 Speedway, Stop C1400, Austin, Texas 78712, USA}

\author{Ely D. Kovetz\,\orcidlink{0000-0001-9256-1144}} 
\affiliation{Department of Physics, Ben-Gurion University of the Negev, Be’er Sheva 84105, Israel}

\author{Marc Kamionkowski\,\orcidlink{0000-0001-7018-2055}}
\affiliation{William H. Miller III Department of Physics and Astronomy, Johns Hopkins University, 3400 N. Charles Street, Baltimore, Maryland, 21218, USA}

\begin{abstract}

We study the phase shift that free-streaming neutrinos imprint on the 21-cm power spectrum during cosmic dawn, computing for the first time its effect on both density- and velocity-induced acoustic oscillations. Neutrinos are known to generate a characteristic phase shift in the acoustic oscillations of the photon–baryon plasma before recombination, a signature already detected in the cosmic microwave background (CMB) as well as the spectrum of baryon acoustic oscillations (BAOs) extracted from galaxy surveys. We show that in the 21-cm signal this phase shift is distinct from that observed in the CMB and BAO spectra, exhibiting a characteristic mode and redshift dependence arising from the additional contribution of the so-called velocity acoustic oscillations (VAOs), sourced by the baryon–dark matter relative velocities. Our results establish the phase of acoustic oscillations in the 21-cm spectrum as a promising new avenue for probing free-streaming light relics at cosmic dawn, complementary to existing CMB and BAO measurements.
\end{abstract}

\date{\today}
\maketitle

\section{Introduction}

\noindent The decoupling of photons from baryons, around the time of hydrogen recombination ($z\sim 1100$), imprinted the acoustic oscillations of the primordial plasma onto cosmological observables. In the cosmic microwave background (CMB), they appear as the familiar pattern of temperature and polarization anisotropies, providing a detailed snapshot of the conditions in the early universe. The same oscillations were frozen into the matter distribution after photon decoupling, and are observed today in the large-scale  structure (LSS) of galaxies ($z\sim0$–$4$), where they are known as baryon acoustic oscillations (BAOs). Together, measurements of these acoustic features in both
the CMB and galaxy surveys form the foundation of modern cosmology, serving as cosmic standard rulers and providing stringent constraints on cosmological parameters~(see e.g.~\cite{Annis:2022xgg, Chang:2022lrw}). 

Beyond these established probes, the redshifted 21-cm line of neutral hydrogen is quickly becoming the next frontier in cosmology, extending our reach into cosmic dawn ($z\sim10$–$30$) when the first stars ignited and began transforming the intergalactic medium (IGM). This signal carries its own acoustic signatures that provide cosmological information complementary to CMB and LSS observations, yet with a crucial distinction. Unlike CMB and galaxy surveys, the 21-cm signal inherits its acoustic structure from two distinct sources: the familiar BAOs arising from matter density fluctuations, and the velocity-induced acoustic oscillations (VAOs), generated by the supersonic relative motion between baryons and dark matter after kinematic decoupling~\citep{visbal12, mcquinn12, fialkov12, Fialkov:2013uwm, munoz19}. These VAOs, which manifest as spatial modulations in early star formation due to streaming velocities, create wavelike features in the 21-cm power spectrum with the same characteristic frequency as BAOs, establishing an additional standard ruler at cosmic dawn~\cite{munoz19b, Sarkar:2022mdz}. While the frequency of these oscillations is a robust cosmological probe of the expansion history of the universe, the relative contributions of BAOs and VAOs change dramatically throughout cosmic dawn~\citep{cruz25},  serving as a sensitive probe of the astrophysical processes governing early star formation and radiative feedback~\citep{munoz20,Libanore:2023oxf,Zhang:2024pwv, verwohlt2024}, as well as any exotic physics that may alter them~\citep{Hotinli:2021vxg,Hotinli:2021xln, cruz24}. For instance, VAOs dominate at early times when streaming velocities suppress star formation in low-mass halos~\citep{mcquinn12, schauer19, kulkarni21}, while BAOs become increasingly prominent at later epochs as density-driven processes take over~\citep{cruz25}. As a whole, this dual acoustic structure makes the 21-cm signal uniquely sensitive to both astrophysics and cosmology, offering novel tests of fundamental physics beyond the reach of CMB and LSS observations alone.

Among the fundamental particles that impact these acoustic oscillations, neutrinos occupy a unique position as the most weakly interacting species in the Standard Model (SM).  Having decoupled from the primordial plasma at temperatures around 1 MeV, roughly one second after the Big Bang, these neutrinos have been freely streaming through the cosmos since, forming the cosmic neutrino background (C$\nu$B), whose energy density is typically parametrized in terms of the effective number of relativistic species~$N_{\rm eff}$. While their feeble interactions with SM particles make direct detection of the C$\nu$B extraordinarily challenging~\cite{Bauer:2022lri,PTOLEMY:2019hkd}, neutrinos contribute a significant fraction of the total radiation density in the early universe, roughly $41\%$ in the standard cosmological model,  leaving sizable and distinctive gravitational imprints on cosmological observables. In addition to their background effect on the expansion rate in the early universe, at the perturbation level neutrinos induce a characteristic phase shift in the acoustic oscillations of the primordial plasma~\citep{Bashinsky:2003tk}. This signature arises from the supersonic propagation of neutrino perturbations after their decoupling. In fact, neutrinos propagate  at nearly the speed of light -- much faster than the sound waves in the tightly coupled photon-baryon fluid -- and as such, they gravitationally pull the plasma ahead of the sound horizon, effectively shifting its acoustic oscillations toward larger scales. The resulting phase shift exhibits a distinctive scale dependence which, under adiabatic initial conditions, can only be produced by free-streaming radiation, making it an exceptionally clean probe of neutrino physics in the early universe~\citep{Baumann:2015rya}. Recent analyses have detected this phase shift at high significance in CMB spectra~\citep{Follin:2015hya,Baumann:2015rya,Montefalcone:2025unv,montefalcone25}, setting strong constraints on the presence of sizable neutrino interactions, which would fundamentally alter this signature~\citep{Choi:2018gho,montefalcone25}. The phase shift has also been identified in LSS data~\citep{Baumann:2017gkg,Baumann:2019keh}, where despite larger uncertainties, it remains robust against nonlinear gravitational evolution~\citep{Baumann:2017lmt}, providing independent evidence for the free-streaming nature of neutrinos.

In this paper, we explore for the first time how the characteristic neutrino-induced phase shift manifests in the 21-cm power spectrum from cosmic dawn. While previous work has established the 21-cm line as a sensitive probe of neutrino properties and light relics~\citep{Lee:2023uxu,Dey:2022ini,Plombat:2024kla,Dhuria:2024zwh,Libanore:2025ack}, the specific signature of the phase shift in this signal has not been investigated. Given the dual acoustic nature of the 21-cm spectrum (with both BAOs and VAOs) we show that accurately extracting the phase information requires a careful modeling of the distinct responses of these oscillatory components to free-streaming neutrinos. 

To this end, we compute for the first time the neutrino-induced phase shift in VAOs and find a clear departure from the well-known BAO response. Specifically, the VAO phase shift is delayed to higher wavenumbers and attains a significantly larger amplitude, yielding a unique phase-shift profile. This difference stems from the fundamentally distinct mathematical origins of BAOs and VAOs, arising from two-point
and four-point functions of baryon density and velocity
fluctuations, respectively.

Since the 21-cm acoustic oscillations are a superposition of these two components~\citep{cruz25}, the net phase creates a unique redshift-dependent signature that interpolates between the VAO and BAO limits, reflecting the evolution of their relative contributions across cosmic dawn. The resulting neutrino-induced phase shift thus provides a new handle on both neutrino physics and astrophysical processes during this epoch, offering a complementary probe with no direct analog in CMB or galaxy survey observations.

This paper is organized as follows. In Sec.~\ref{sec:2}, we review how free-streaming neutrinos induce a characteristic phase shift in the acoustic oscillations of the primordial plasma. In Sec.~\ref{sec:3},  we introduce the basics of the 21-cm signal and discuss its dual acoustic structure  from BAOs and VAOs. In Sec.~\ref{sec:phase_21cm}, we derive the neutrino-induced phase shift in the 21-cm power spectrum, first establishing the phase-shift templates for both BAO and VAO spectra, and then demonstrating how their weighted superposition creates a unique redshift-dependent signature. We conclude in Sec.~\ref{sec:5}.

Throughout this work, we compute the 21-cm spectra with the updated \texttt{zeus21}~\cite{munoz23,munoz23b,cruz24,cruz25} code, which calculates the brightness temperature using baryon fluctuations rather than total-matter fluctuations, in line with the fact that the signal traces the IGM gas~\cite{Furlanetto:2006jb, Pritchard:2008da, Pritchard:2010pa, pritchard12,Flitter:2024eay}. This choice improves the modeling of BAO and VAO features by avoiding the small but non-negligible differences introduced when assuming baryons perfectly trace the total matter~\cite{cruz25}. Additionally, we adopt a {\it Planck 2018}–inspired fiducial $\Lambda$CDM model (Table~\ref{tab:parameters}) and fix all astrophysical parameters to the \texttt{zeus21} default choices~\cite{cruz24}, unless otherwise stated.

\section{Free-streaming Neutrinos and the Phase Shift}
\label{sec:2}
\begin{table}
	\centering
	\sisetup{group-digits=false}
	\begin{tabular}{l S[table-format=1.5]}
			\toprule
		Parameter 				& {Fiducial Value}															\\
			\midrule[0.065em]
		$\omega_b$ 				& 0.02238 				 		\\
		$\omega_c$				& 0.12011					\\
		$100\,\theta_s$ 		& 1.04178 						\\
		$\ln(\num{e10}A_s)$		& 3.0448							\\
		$n_s$					& 0.96605 													\\
		$\tau$ 					& 0.0543 									\\
		$N_{\rm eff}$					& 3.044 							\\
		$Y_p$					& {`BBN'\hskip6.5pt}\\
    
			\bottomrule
	\end{tabular}
	\caption{Fiducial $\Lambda$CDM parameters based on the {\it Planck 2018} best-fit cosmology~\cite{Planck:2018vyg}. Here, $\omega_b$ and $\omega_c$ are the physical baryon and cold dark matter densities, respectively; $\theta_s$ is the angular size of the sound horizon at decoupling; $\ln(10^{10}A_s)$ and $n_s$ are respectively the logarithm of the primordial scalar spectrum amplitude and its spectral index at the pivot scale $k_\star = 0.05\,{\rm Mpc}^{-1}$; $\tau$ is the optical depth due to reionization; $N_{\rm eff}$ is the effective number of free-streaming relativistic species; and $Y_p$ is the primordial helium fraction. Neutrinos are treated as massless throughout, which is a valid approximation for the purpose of the phase-shift extraction~\cite{Montefalcone:2025unv}. The primordial helium abundance is generally fixed by requiring consistency with Big Bang Nucleosynthesis for a given choice of $(\omega_b, \,N_{\rm eff})$, yielding $Y_p = 0.24534$ for this set of parameters.}
\label{tab:parameters}
\end{table}
\noindent Neutrinos, as the most weakly coupled particles in the SM, are the first known species to decouple from the primordial plasma at redshift $z_{\nu,\rm dec}\sim 10^{10}$. Since then, neutrinos have been freely streaming through the universe at nearly the speed of light, influencing the evolution of the photon-baryon fluid, which remains coupled for much longer, until hydrogen recombination is complete at $z_{\rm rec}\approx z_{\gamma,\rm dec}\approx1080$. In fact, the propagation of neutrino perturbations at speeds much faster than the sound speed of the primordial plasma modifies the time-dependent gravitational driving of the acoustic oscillations in the photon-baryon fluid, inducing both an amplitude and a phase shift~\cite{Bashinsky:2003tk,Baumann:2015rya}. Effectively, due to their large free-streaming length, neutrinos  overtake the primordial plasma and gravitationally pull it towards larger scales, ahead of the sound horizon. Schematically, the evolution of the density fluctuations in the photon-baryon fluid, $\delta_{\gamma,b}$, at wavenumber $k=|\vec{k}|$, are altered as:
\begin{equation}
    \delta_{\gamma,b}(\vec{k}) \approx B_{\gamma,b}(\vec{k}) \cos(k r_s) \to \tilde{B}_{\gamma,b}(\vec{k}) \cos(k r_s + \phi)\,\label{eq:phase0}
\end{equation}
where we implicitly assumed adiabatic initial conditions, $r_s$ is the size of the sound horizon, and $\phi$ and $B_{\gamma,b} \to \tilde{B}_{\gamma,b}$ capture the corresponding phase and amplitude shift induced by free-streaming neutrinos. 

While the induced amplitude shift exhibits a smooth dependence on angular scales that can be trivially mimicked by varying initial conditions, the phase shift represents a robust signature of free-streaming neutrinos~\citep{Bashinsky:2003tk,Baumann:2015rya}. In particular, the induced phase shift approaches a constant for modes that enter the horizon deep in radiation domination, while it acquires a characteristic wavenumber dependence which smoothly goes to zero for modes that enter the horizon during matter domination, when 
radiation becomes gravitationally subdominant~\cite{Baumann:2015rya}. Most importantly, under adiabatic initial conditions, such a coherent shift in the acoustic oscillations of the primordial plasma can only be produced by free-streaming radiation~\cite{Baumann:2015rya}, making the phase shift a pristine signal to look for in cosmological observables. Specifically, this signature is imprinted in CMB spectra, as well as the BAO spectra extracted from large-scale structure (LSS) surveys (see Fig.~\ref{fig:1}). Dedicated analyses targeting this effect have already provided strong evidence for neutrino free-streaming, in turn placing tight constraints on potential neutrino interactions~\cite{Follin:2015hya,Montefalcone:2025unv,Choi:2018gho,Whitford:2024ecj,Saravanan:2025cyi,montefalcone25}.

Evidently, the neutrino-induced phase shift will also appear in the acoustic oscillations of the 21-cm line spectrum, which are sourced by both baryon density and velocity fluctuations~\citep*[see also \citep{barkana06, mao12}]{munoz19, munoz19b, Cruz:2024fsv}. The derivation of this effect forms the main subject of our work and results in a unique signature, distinct from the imprint in CMB and BAO observations. As we will show in detail in Sec.~\ref{sec:phase_21cm}, this distinction ultimately arises from the presence of VAOs, which by rotational invariance can only depend on the square of the relative baryon-DM velocities, effectively making them four point functions of the underlying velocity perturbations that inherit a fundamentally different neutrino-induced phase shift than the two-point density correlations driving standard BAOs.

\section{The 21-cm Signal and its Acoustic Features}\label{sec:3}

In this section, we review the fundamentals of the 21-cm signal and demonstrate how it inherits a rich acoustic structure from both BAOs and VAOs. We establish the framework for isolating these contributions and discuss their potential for probing fundamental cosmology and astrophysics, through both their amplitude, frequency and phase information.
\subsection{The Basics of the 21-cm Signal}
\noindent Neutral hydrogen atoms along the line of sight to the CMB can undergo a spin-flip transition that creates either a 21-cm emission or absorption line, which gets redshifted as the universe expands. Observations of the redshifted 21-cm line probe various epochs, in particular from cosmic dawn to reionization, providing a unique window into high-redshift cosmology and the astrophysical processes governing the formation of the first stars and galaxies. The 21-cm signal is usually quantified in terms of the brightness temperature~\citep{Furlanetto:2006jb}:
\begin{equation} \label{eq:T21}
    T_{21}=T_0(z)\left(1+\delta_b-\delta_v\right)\, x_{\mathrm{HI}}\left(\frac{x_\alpha}{1+x_\alpha}\right)\left(1-\frac{T_{\mathrm{CMB}}}{T_c}\right),
\end{equation}
where  $T_0(z)$ is simply a normalization factor that depends on the cosmic baryon and total matter densities, $x_{\rm HI}$ is the neutral hydrogen fraction, $x_\alpha$ is the WF parameter which measures the Lyman-$\alpha$ flux from early galaxies, and the final factor contrasts the CMB temperature $T_{\rm CMB}$ with the gas color temperature $T_c$, determined by the balance between adiabatic cooling and X-ray heating. In addition, this expression leverages a few standard approximations following~\citep{munoz23,munoz23b,Cruz:2024fsv}; namely we ignore collisional coupling, use linear-order redshift-space distortions, and work in the limit of small optical depth $\tau_{21} \ll 1$. Under these conditions, the equation cleanly separates cosmological effects -- encoded in the local baryonic density contrast $\delta_b$ and bulk velocity $\delta_v$ --
from astrophysical processes governing reionization, Wouthuysen-Field (WF) coupling, and IGM heating. 

Beyond the global signal, the spatial anisotropies $\delta T_{21}$ in the brightness temperature field trace the inhomogeneous growth of structure and radiative backgrounds, with their statistics encapsulated to first order through the 21-cm power spectrum $P_{21}(k, z)$:
\begin{equation}
    \langle \delta T_{21}(\vec{k}, z)\delta T^*_{21}(\vec{k}^\prime,z)\rangle= (2\pi)^3\delta_D(\vec{k} + \vec{k}^\prime) P_{21}(\vec{k}, z).
\end{equation}

The standard thermal history of the 21-cm signal during cosmic dawn proceeds as follows. At early times ($z \sim 25-30$), the formation of the first Pop III stars produces UV radiation that couples the hydrogen spin temperature to the cold gas temperature through resonant Lyman-$\alpha$ scattering, i.e. the WF effect~\citep{wouthuysen52,field59,hirata06}.

\begin{figure*}
    \centering
    \includegraphics[width=1\linewidth]{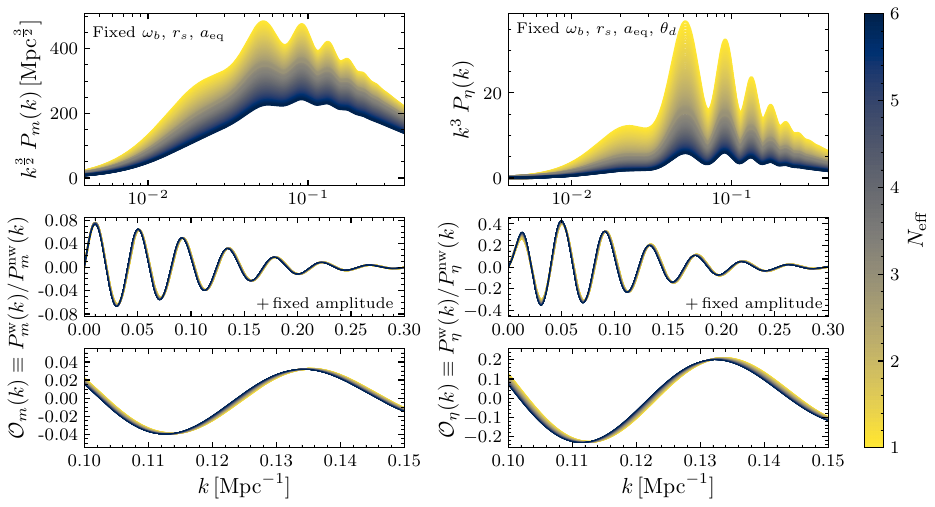}
    \caption{ Illustration of the phase shift as it appears in the matter power spectrum $P_m(k)$ ({\bf Top left}) and the BAO spectrum, $\mathcal{O}_m(k)\equiv P_m^{\rm w}(k)/P_m^{\rm nw}(k)$, ({\bf Bottom left}) as a function of the effective number of relativistic species $N_{\rm eff}$. Similarly for the scalar power spectrum of the baryon-DM relative velocity, $P_\eta(k)$, ({\bf Top right}) and the VAO spectrum, $\mathcal{O}_\eta(k)\equiv P_\eta^{\rm w}(k)/P_\eta^{\rm nw}(k)$ ({\bf Bottom right}). To isolate the phase shift, we keep the following quantities fixed across all panels as done in~\cite{Follin:2015hya,Baumann:2017gkg,Montefalcone:2025unv}: the physical baryon density $\omega_b$, the scale factor at matter-radiation equality $a_{\rm eq}$ and the physical sound horizon at the drag epoch $r_s$. For the velocity spectra, we additionally fix the angular damping scale $\theta_d$ by adjusting the primordial helium fraction. The full spectra in the top panels are rescaled by $k^{3/2}$ and $k^3$, respectively, to reduce the range of scales. The oscillatory spectra in the lower four panels are normalized at the fourth peak, illustrating the remaining phase shift. Finally, bottom panels provide further zoom in around a narrow $k$-range to highlight the phase shift more clearly.}
    \label{fig:1}
\end{figure*}
 Since the gas has been cooling adiabatically and is much colder than the CMB, this Lyman-$\alpha$ coupling epoch (LCE) produces strong 21-cm absorption. As cosmic dawn progresses, more stars form and die, and their remnants begin producing copious X-rays that heat the IGM. The competition between efficient Lyman-$\alpha$ coupling and the onset of heating drives the global 21-cm absorption trough, which in our fiducial model occurs near $z \sim 19$. Thereafter, during the so called epoch of heating (EoH), X-rays continue to raise the gas temperature until it exceeds that of the CMB, causing the 21-cm signal to transition from absorption to emission, around $z \sim 13$ in our model. Throughout these transitions, spatial variations in star formation -- modulated by both density perturbations and streaming velocities -- imprint BAOs and VAOs onto the 21-cm power spectrum, producing the acoustic features we examine in the following section.

\subsection{Acoustic Oscillations in the 21-cm Power Spectrum}

\noindent The 21-cm power spectrum exhibits a rich acoustic structure, inherited from its underlying constituents. As evident from Eq.~\eqref{eq:T21}, the brightness temperature is sourced by the fractional baryonic density $\delta_b$~\citep{Furlanetto:2006jb, Pritchard:2008da, Pritchard:2010pa, pritchard12}, as well as the total matter density contrast $\delta_m$ through the halo mass function and star formation rate density that depend on the gravitational collapse of all matter components~\citep{Cruz:2024fsv}. The baryonic density field contains the familiar BAOs -- oscillatory features imprinted by sound waves in the pre-recombination photon-baryon plasma that froze out at the drag epoch -- that we observe today in the CMB and large-scale structure surveys. In the context of the 21-cm spectrum, these density-sourced BAOs modulate the signal through their direct influence on the neutral hydrogen distribution and indirectly through their effect on star formation rates in overdense regions~\citep{tseliakhovich10, dalal10}.  To isolate these BAO contributions, we split the baryon ($P_b$), total matter ($P_m$), and baryon-matter cross-correlation ($P_{bm}$) power spectra into a smooth (`no-wiggle') and an oscillatory (`wiggle') component~(see Fig.~\ref{fig:1}), namely:
\begin{equation}
    \mathcal{O}_{j,\rm BAO}(k)\equiv P_{j}^{\rm w}(k)/P_{j}^{\rm nw}(k)
\end{equation}
where $j=\{b,m,bm\}$ depending on the source term under consideration. To perform this separation, we compute $P_j^{\rm nw}(k)$ by sine transforming the power spectra to discrete real space where the acoustic oscillations map to a localized bump, remove this feature, and inverse transform back to Fourier space, following the established methodology of~\cite{Hamann:2010pw,Baumann:2017gkg}. The `wiggle' component of the power spectrum is then simply defined as the difference between the full spectrum and the `no-wiggle' component, i.e. $P_j^{\rm w}(k)\equiv P_j(k)-P^{\rm nw}_j(k)$.

The same BAOs that created the density oscillations also generated supersonic relative velocities between baryons and dark matter, $\vec{v}_{\rm cb}$, with amplitudes of approximately $\sim 30\,$km/s at recombination~\citep{tseliakhovich10, dalal10}. These streaming velocities suppress star formation in low-mass halos by reducing small-scale structure formation~\citep{mcquinn12,fialkov12,tseliakhovich10,naoz12, bovy13,munoz19}, disrupting protogalactic cooling~\citep{dalal10, greif11, hirano18, schauer19}, and inhibiting gas accretion~\citep{dalal10, tseliakhovich11, stacy11, oleary12, naoz13}.  Since the relative velocity field inherits the acoustic correlations from the original BAOs, the spatial distribution of these velocities, which in turn dictates the spatial distribution of early star formation, imprints a second set of oscillatory features in the 21-cm power spectrum, known as VAOs~\citep{visbal12, mcquinn12, fialkov12, Fialkov:2013uwm, munoz19}.  Recent work has shown that VAO contributions are simply exponential correlations of the scalar power spectrum of the relative velocity $P_\eta(k)$~\cite{Cruz:2024fsv} which in configuration space 
can be expressed as~\cite{dalal10}:
\begin{equation}
    P_\eta(k) =4 \pi \int d r r^2 j_0(k r)\left[6 \psi_0(r)^2+3 \psi_2(r)^2\right], \label{eq:Peta}
\end{equation}
with the scalar field $\eta$ defined according to:
\begin{equation}
    \eta(\vec{r})\equiv \frac{v^2_{\rm cb}(\vec{r},z)}{\left\langle v_{\rm cb}^2(z)\right\rangle}-1,
\end{equation}
which is normalized to have zero mean and, to excellent approximation, is redshift independent~\cite{ferraro12}. The correlation functions $\psi_0(r),\,\psi_1(r)$ in Eq.~\eqref{eq:Peta} are given by:
\begin{align}
    \psi_0(r)&=\frac{1}{3\sigma_{\rm cb}^2} \int \frac{k^2 d k}{2 \pi^2} P_{v_{\mathrm{cb}}}(k) j_0(k r) \\
    \psi_2(r)&=-\frac{2}{3\sigma_{\rm cb}^2} \int \frac{k^2 d k}{2 \pi^2} P_{v_{\mathrm{cb}}}(k) j_2(k r) 
\end{align}
in terms of spherical Bessel functions $j_\ell(x)$, the root-mean-square of the relative velocity $\sigma_{\rm cb}\equiv\sqrt{\left\langle v_{\rm cb}^2\right\rangle}$, and its power spectrum $P_{v_{\rm cb}}$:
\begin{equation}
P_{v_{\mathrm{cb}}}(k)=A_s\left(\frac{k}{k_{\mathrm{pivot}}}\right)^{n_s-1}\left[\frac{\theta_b(k)-\theta_c(k)}{k}\right]^2 \frac{2 \pi^2}{k^3}.\label{eq:Pvcb}
\end{equation}
Here $A_s$ and $n_s$ are the amplitude and spectral index of the primordial power spectrum, $k_{\rm pivot}=0.05\,\rm{Mpc}^{-1}$ and $\theta_b$ and $\theta_c$ are the baryon and cold dark matter transfer functions, respectively. While the power spectrum $P_{v_{\rm cb}}(k)$ describes the fluctuations in the vector velocity field itself, the relevant quantity for physical observables is the power spectrum $P_\eta(k)$ which is effectively a four-point function in $v_{\rm cb}$, reflecting the fact that observable quantities can only depend on $v_{cb}^2$ by rotational invariance. In addition, while early models assumed that the imprint from the VAOs was fixed at kinematic decoupling $z_{\rm kin}\approx 1060$, the scalar velocity spectrum $P_\eta(k,z)$ is only roughly redshift-independent for $z\lesssim 100$~\citep{ferraro12, Yoo:2011tq}. To model accurately the velocity-modulated signatures in the IGM, we therefore evaluate the velocity transfer function closer to the dark ages, specifically at $z=50$, rather than at kinematic decoupling~\cite{cruz25}. Hydrodynamical simulations confirm that relative velocities suppress star formation through a redshift-independent power-law increase in the molecular cooling mass threshold~\citep{schauer19, kulkarni21, munoz22}, validating this parametrization for modeling velocity effects during cosmic dawn.

As done for the BAO spectrum, we can isolate the VAO contribution by splitting the $\eta$ power spectrum into its smooth and oscillatory component~(see Fig.~\ref{fig:1}), such that:
\begin{equation}
    \mathcal{O}_{\eta,\rm VAO}(k)\equiv P^{\rm w}_\eta(k)/P^{\rm nw}_\eta(k).
\end{equation}
The total acoustic signal in the 21-cm power spectrum can then be decomposed as:
\begin{equation}
    P_{21}(k,z)= P_{21}^{\rm nw}(k,z)\left[1+\mathcal{O}_{21,\rm tot}(k,z)\right] \label{eq:O21_v0}
\end{equation}
where $P^{\rm nw}_{21}(k,z)$ is the 21-cm power spectrum sourced only by the `no-wiggle' components of the density and relative-velocity power spectra, and the oscillatory component:
\begin{equation}
    \mathcal{O}_{21,\rm tot}(k,z) \approx \mathcal{O}_{21,\rm BAO}(k,z)+\mathcal{O}_{21,\rm VAO}(k,z) \label{eq:O21}
\end{equation}
is to an excellent approximation just the sum of its BAO and VAO contributions. This simplification is due to the minimal correlation between VAOs and the usual 21-cm fluctuations sourced by matter and baryon overdensities~\cite{dalal10,alihaimoud14, Munoz:2018jwq}. The individual BAO and VAO oscillatory contributions are obtained  by computing the 21-cm power spectrum with different combinations of the source spectra, namely using the total baryon and matter auto- and cross-correlation spectra with the no-wiggle $\eta$ spectrum for $\mathcal{O}_{21,\rm BAO}(k,z)$, while $\mathcal{O}_{21,\rm VAO}(k,z)$ uses the full $\eta$ spectrum with the no-wiggle density spectra~\cite{cruz25}.

Crucially, the acoustic oscillations in the source spectra $\mathcal{O}_{j,\rm BAO}$ and $\mathcal{O}_{\eta,\rm VAO}$ are determined by the underlying cosmology, reflecting the oscillations in the photon-baryon plasma that froze out after recombination at the characteristic sound horizon scale $r_s$. This makes both BAOs and VAOs peak positions in the 21-cm spectrum, inherited from the corresponding acoustic signals in the source spectra, robust standard rulers independent of the complex astrophysics during cosmic dawn~\cite{wyithe08, wyithe08b, loeb08, chang08, seo10, munoz19b, Sarkar:2022mdz}. In addition, the relative amplitudes of the BAOs and VAOs contribution to the 21-cm signal exhibit a strong redshift dependence, reflecting the evolution of the underlying astrophysical processes~\cite{munoz20,Hotinli:2021vxg,Libanore:2023oxf, cruz24,Zhang:2024pwv, verwohlt2024}. At early times, during the LCE ($z\gtrsim 25$), VAOs dominate, as streaming velocities effectively suppress star formation in the smallest dark-matter halos. Later, during the EoH ($z\sim 17$) the VAO contribution peaks again as larger velocities reduce the number of X-ray photons produced, leading to cooler gas and a more negative $T_{21}$. During the transition between these two epochs ($z\sim 19$), the VAO amplitude nearly vanishes as the opposing influences on $T_{21}$, from the production of fewer Lyman-$\alpha$ and X-ray photons, respectively, balances out. In contrast, BAOs become increasingly prominent at intermediate redshifts ($z\sim 13-17$) as accreting halos grow immune to velocity suppression effects and density-driven star formation takes over. In addition to this complex redshift dependence, there is also a scale dependence, with BAOs generically dominating on smaller scales, $k\gtrsim 0.3\,{\rm Mpc}^{-1}$, primarily due to the intrinsically smaller amplitude of the source velocity spectrum $P_\eta(k)$ relative to the matter spectrum $P_m(k)$ on small scales~\citep{cruz25}. The VAOs are also further damped from the non-locality of photon propagation~\cite{munoz19,Cruz:2024fsv}. In fact, during cosmic dawn, Lyman-$\alpha$ photons travel distances of $\sim 10^2\,$kpc while X-rays travel $\sim 1-10\,$Mpc. As cosmic time progresses, these photons travel increasingly longer distances, causing the VAO signal to become progressively more washed out at small scales as the extended photon propagation smooths out velocity-induced fluctuations.

This dual acoustic structure provides complementary cosmological and astrophysical information: the peak positions, inherited from the primordial oscillations of the photon-baryon plasma, constrain fundamental cosmological parameters through the sound horizon, while the amplitude evolution and relative contributions of BAOs versus VAOs encode the detailed physics of early star formation, radiative feedback, and IGM heating during cosmic dawn. We will explicitly show how these contributions shape the 21-cm signal in more detail in Sec.~\ref{sec:phase_21cm}.

Beyond the amplitude information, the phase of the BAO and VAO oscillations also constitutes a robust probe of fundamental cosmology, encoding recombination-era physics that remains protected from gravitational nonlinearities~\cite{Pan:2016zla,Baumann:2017lmt,Green:2020fjb}. In addition, while both BAOs and VAOs oscillate at the same characteristic frequency set by the sound horizon $r_s$, they are not expected to exhibit the same phase, as imprinted in the acoustic oscillations of the photon-baryon plasma. This is due to the fundamental distinction between the $\eta$ spectrum and the baryon and matter auto- and cross-correlation spectra: the former represents a four-point function in the underlying velocity perturbations while the latter are two-point functions of their respective density perturbations, leading to different responses to early-universe physics. In the following section, we will demonstrate how the distinct mathematical structure of BAOs and VAOs produces a unique mode and redshift-dependent signature in the 21-cm spectrum, complementary to the phase shift observed in the CMB and large-scale structure.

\section{Extracting the Phase Shift in the 21-cm power spectrum}\label{sec:phase_21cm}

Having established the dual acoustic structure of the 21-cm power spectrum, we now turn to extracting the neutrino-induced phase shift from its BAO and VAO components. We begin by reviewing how the neutrino phase shift manifests in the standard BAO spectrum, establishing the baseline expectation from density-sourced oscillations. We then derive the corresponding phase shift in the VAO spectrum, which constitutes the primary theoretical contribution of this work, showing how the four-point nature of velocity correlations leads to a modified response to neutrino perturbations. Finally, we combine these results to demonstrate how the interplay between BAO and VAO phase shifts produces a distinctive redshift-dependent signature in the 21-cm power spectrum that is unique compared to other cosmological probes.

\subsection{The Phase Shift in the BAOs}
\noindent The phase shift in the BAO spectrum induced by $N_{\rm eff}$ free-streaming relativistic species can be parametrized as~\cite{Baumann:2017gkg,Baumann:2019keh,Green:2020fjb}:
\begin{equation}
    \phi(k; N_{\rm eff})= \beta(N_{\rm eff})f_\phi(k), \label{eq:dphi}
\end{equation}
with:
\begin{align}
    \beta(N_{\rm eff})&\equiv \frac{\epsilon(N_{\rm eff})}{\epsilon(3.044)},\label{eq:beta} \\
    f_\phi(k)&=\frac{\phi_{\infty}}{1+(k_\star/k)^\xi},\label{eq:fphi}
\end{align}
 where $\epsilon(N_{\rm eff})\equiv \rho_\nu/\rho_r=N_{\rm eff}/\left(a_\nu +N_{\rm eff}\right)$ represents the fractional neutrino energy density relative to the total radiation density $\rho_r=\rho_\gamma+\rho_\nu$, with $a_\nu\equiv 8/7\left(11/4\right)^{4/3}\approx 4.40$. The amplitude $\beta(N_{\rm eff})$ characterizes the overall size of the effect. It is proportional to the fractional energy density of neutrinos~\cite{Bashinsky:2003tk, Baumann:2015rya}, and  normalized to the shift between the Standard Model value of $N_{\rm eff}^{\rm SM}=3.044$ and zero neutrino species, following previous studies~\cite{Baumann:2017gkg,Baumann:2019keh}.
 On the other hand, the template function $f_\phi(k)$ fully encodes the mode dependence of the induced phase shift, with its functional form reflecting the physical timescales governing neutrinos effects on the primordial acoustic oscillations. In particular, $f_\phi(k)$ asymptotically approaches a constant $\phi_\infty$ for $k\rightarrow\infty$, matching the expected behavior in a radiation-dominated universe~\cite{Bashinsky:2003tk,Baumann:2015rya}.  Conversely, at large scales, as we move  from modes that entered deep inside radiation-domination into the matter-dominated epoch, the relative contribution of neutrinos to the total energy density decreases, which results in a smooth suppression of the phase which vanishes to zero at least as $\sim k^2$ for $k \to 0$~\cite{Baumann:2015rya}. This characteristic wavenumber dependence at large scales is governed in $f_\phi(k)$ by the parameters $k_\star$ and $\xi<0$. As
illustrated in Fig.~\ref{fig:3}, this fitting function well approximates the shifts in the zeros, peaks and troughs of the BAO spectrum with the best-fit parameters listed in Table~\ref{tab:phase_templates}, matching well the established fit from the literature~\cite{Baumann:2017gkg}. 

To obtain this phase-shift template, we closely follow the methodology in~\cite{Baumann:2017gkg}. In summary, we numerically calculated the mode shifts $\delta k$ from the BAO spectra obtained using the Boltzmann solver \texttt{CLASS}~\cite{Blas:2011rf}, comparing our fiducial $\Lambda$CDM model against 100 cosmologies with varying  $N_{\rm eff} \in [0, 3.3]$.\footnote{Following~\cite{Baumann:2017gkg,Baumann:2019keh}, we restrict to $N_{\rm eff} \in [0,3.3]$ due to a small, unexpected feature in the extracted peak locations near $N_{\rm eff} \simeq 3.33$, with smooth variation on either side. We find a similar behavior in our analysis, and so we adopt the same range for consistency, noting that our main interest lies in the smaller $N_{\rm eff}$ values, and that the resulting template should remain valid beyond this interval} To isolate the effect from the free-streaming of neutrinos, we kept fixed across all models the physical baryon density $\omega_b$ and the scale factor at matter-radiation
equality $a_{\rm eq}$. We also rescaled all wavenumbers as $k \to r_s^{\rm fid}/r_s \cdot k$, to match the sound horizon at the drag epoch to its fiducial value, thereby removing the frequency shift in the BAO pattern caused by varying $N_{\rm eff}$. Finally, we undamped the BAO spectra to facilitate the extraction of the zeros, peaks and troughs locations, and normalized the amplitude of the fourth peak across all models for convenience,  equivalently to
what we showed in the left panel of Fig.~\ref{fig:1}.

Lastly, while derived from the matter power spectrum, the same phase-shift template equally applies to the BAOs in the baryon auto- and baryon-matter cross-correlation spectra, since the neutrino-induced phase is fundamentally imprinted in the underlying baryon density fluctuations, as schematically illustrated in Eq.~\eqref{eq:phase0}.
\subsection{The Phase Shift in the VAOs}\label{sec:phase_vao}

\begin{table}
	\centering
	\sisetup{group-digits=false}
	\begin{tabular}{l c @{\hskip 1em} c}
			\toprule
		Parameter 				& {BAO Template} & 							{VAO Template}								\\
			\midrule[0.065em]
		$\phi_\infty$ 				& $0.27 \pm 0.03$ & $0.36 \pm 0.02$				 		\\
		$k_\star\,\left[\rm{Mpc}^{-1}\right]$				& $0.04 \pm 0.01$ &	 $0.09 \pm 0.01$				\\
		$\xi$ 		& $0.74\pm 0.18$ & $1.40\pm 0.08$ \\

			\bottomrule
	\end{tabular}
	\caption{Best-fit parameters of the phase-shift template $f_\phi(k)$, Eq.~\eqref{eq:fphi}, for both the BAO and VAO spectra. The parameter $\phi_\infty$ sets the asymptotic phase shift at large $k$, while $k_\star$ and $\xi$ govern the scale dependence of the transition to this limit. The BAO fit is consistent with previous results~\cite{Baumann:2017gkg}, while the VAO fit shows a markedly different shape, characterized by a larger asymptotic amplitude and smoother transition scale.}
\label{tab:phase_templates}
\end{table}

\noindent Building on the physical connection between density and velocity perturbations, a naive expectation would be that the phase shift in the VAOs mirrors that in the BAOs. In fact, VAOs ultimately originate from the baryon velocity divergence $\theta_b$, which is related to the baryon density perturbations effectively by a total time derivative, thus experiencing the same neutrino-induced phase shift prior to the drag epoch~\cite{Green:2020fjb, Montefalcone:2025unv}. Indeed, this expectation holds for the power spectrum of the relative velocity vector field $P_{v_{\rm cb}}(k)$, which represents a two-point correlation function of the baryon velocity divergence. 

\begin{figure}
    \centering
    \includegraphics[width=\linewidth]{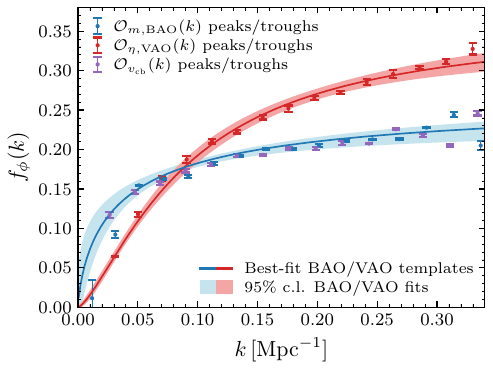}
    \caption{Templates of the BAO and VAO phase shift $f_\phi(k)$ as defined in Eq.~\eqref{eq:dphi}, as a function of wavenumber $k$. The numerical
phase shifts for the BAO and VAO spectra, displayed in \textcolor{tabblue}{blue} and \textcolor{tabred}{red}, respectively, were obtained from sampling 100 different cosmologies with varying number of relativistic species $N_{\rm eff}$ and rescaled by $\beta(N_{\rm eff})$, as defined in Eq.~\eqref{eq:beta}. The error
bars indicate the standard deviation in these measurements at the respective peaks/troughs relative to the fiducial cosmology. Solid lines represent the best-fit BAO and VAO templates, with shaded regions illustrating the corresponding $2\sigma$ confidence intervals. For reference, numerical phase shifts derived from the acoustic oscillations in the relative-velocity power spectrum, $\mathcal{O}_{v_{\rm cb}}(k)\equiv P^{\rm w}_{v_{\rm cb}}(k)/P^{\rm nw}_{v_{\rm cb}}(k)$, are also shown in \textcolor{tabpurple}{purple}.}
    \label{fig:3}
\end{figure}

We confirm this by applying the same methodology outlined in the previous section to extract the phase shift from $P_{v_{\rm cb}}(k)$, with one additional correction: we adjust the primordial helium fraction $Y_p$ to fix the damping scale $\theta_d$ across all cosmologies with varying $N_{\rm eff}$, as done in the context of the CMB phase shift extraction~\cite{Follin:2015hya, Montefalcone:2025unv}. While varying or keeping $Y_p$ fixed has no effect on the phase shift extraction from baryon density perturbations, it significantly helps isolate the phase shift originating from $\theta_b$ due to its more direct connection to photon perturbations, which are strongly affected by diffusion damping, and directly source velocity perturbations via Compton scattering with electrons. Figure~\ref{fig:3} explicitly demonstrates that the phase shift in $P_{v_{\rm cb}}$ matches that found in the BAO spectrum. The extraction of the phase shift from the oscillatory features of $P_{v_{\rm cb}}$ is further illustrated in Figure~\ref{fig:A1} in the Appendix, analogous to left panel of Fig.~\ref{fig:1} displaying the BAO spectrum from the matter power spectrum. 

While this direct correspondence holds for $P_{v_{\rm cb}}(k)$, as described in detail in the previous section, the relevant source term for the VAO features imprinted in the 21-cm power spectrum is the scalar velocity power spectrum $P_\eta(k)$, which effectively represents a four-point function in the relative velocity field rather than a simple two-point correlation. Given this fundamental mathematical distinction, we expect a different response to the neutrino-induced phase shift in the VAO spectrum, which is still well parametrized by the same template function as in Eq.~\eqref{eq:dphi},  now with VAO-specific best-fit parameters. To understand how this phase shift manifests in $P_\eta(k)$, it is useful to rewrite it in its Fourier space representation~\cite{ferraro12}:
\begin{align}
    P_\eta(k)&=  \frac{1}{4\pi^2k\,\sigma_{\rm cb}^2}\int_0^\infty dk_1 \int_{\max(k_1, |k-k_1|)}^{k+k_1} dk_2  \nonumber \\
    &\times \frac{(k^2 - k_1^2 - k_2^2)^2}{k_1 k_2} P_{v_\mathrm{cb}}(k_1)P_{v_\mathrm{cb}}(k_2)
\end{align}
This expression explicitly illustrates that each mode $k$ in the $\eta$ spectrum receives contributions from all possible pairs of velocity modes $(k_1,k_2)$ that can combine to form $k$,  weighted by a characteristic geometric factor. The weighting $\left(k^2 - k_1^2 -k_2^2\right)^2/\left(k_1 k_2\right) = k_1 k_2\cos^2(\varphi)$, where $\varphi$ is the angle between $\vec{k}_1$
and $\vec{k}_2$, strongly favors aligned configurations while suppressing perpendicular ones, which is a consequence of the scalar nature of $\eta \propto v_{\rm cb}^2$. 

This four-point structure leads to three key predictions for the VAO phase shift. First, in the limit of $k\rightarrow \infty$, the integral becomes dominated by nearly equal modes $k_1 \approx k_2 \approx k/2$ in aligned configurations. Since both modes carry the  same phase shift $\phi_{\infty, \rm BAO}$ from the source $P_{v_{\rm cb}}$ spectrum, and $P_\eta$ is constructed from their product $P_{v_{\rm cb}}(k_1)P_{v_{\rm cb}}(k_2)$, the phases add linearly to yield an asymptotic phase shift $\sim 2\phi_{\infty,\rm BAO}$. 
Second, the transition scale $k_{\star,\rm VAO}$ will be larger than the BAO value, with $k_{\star,\rm VAO} \gtrsim 2k_{\star, \rm BAO}$. In fact, the induced phase shift in the VAOs will exceed the phase sourced in $P_{v_{\rm cb}}$ when both contributing modes satisfy $k_1, k_2 \gtrsim k_{\star, \rm BAO}$, which for symmetric configurations requires $k \gtrsim 2k_{\star, \rm BAO}$. Since the transition scale where the VAO phase shift saturates occurs at or beyond this threshold, we expect $k_{\star,\rm VAO} \gtrsim 2 k_{\star, \rm BAO}$, with the factor of two arising in the limit of a step-like phase shift in the source spectrum. Finally, we also expect the steepness parameter $\xi$ to increase in magnitude due to the mode-mixing inherent in the four-point structure of $P_\eta$ which averages out the sharp scale dependence present in the original phase imprinted in the relative velocity two-point spectrum, resulting in a smoother transition between large and small scales. 

The right panels of Figure~\ref{fig:1} provide a visual representation of the neutrino-induced phase shift in the $\eta$ spectrum. Following the equivalent extraction procedure used for $P_{v_{\rm cb}}$, we find that the fitting function $f_\phi(k)$ well describes the phase shift in the VAO spectrum with the best-fit parameters given in Table~\ref{tab:phase_templates}. This VAO phase-shift template, summarized in Fig.~\ref{fig:3}, represents one of our main results -- establishing for the first time the characteristic neutrino-induced phase signature in velocity-driven acoustic oscillations. These results broadly confirm our analytic expectations, though with some notable deviations. The asymptotic phase shift $\phi_{\infty,\rm VAO} \approx 0.36$ is somewhat smaller than the predicted factor of two relative to $\phi_{\infty,\rm BAO} \approx 0.27$, yielding a ratio of $\sim 1.5$ instead. Similarly, the transition scale $k_{\star,\rm VAO} \sim 3 k_{\star,\rm BAO}$ exceeds our expectation of doubling. These discrepancies arise from the presence of the no-wiggle component in the source $P_{v_{\rm cb}}$ spectrum. As we outlined above, the four-point integral effectively weights different mode combinations based on both the geometric factor and the amplitude of the source spectra. Since the no-wiggle component lacks phase information, mode pairs involving this smooth background dilute the coherent phase addition we assumed in our simplified analysis. In this sense, there is only a subset of mode pairs where both $k_1$ and $k_2$ are sourced predominantly by the oscillatory contribution in $P_{v_{\rm cb}}$, which explains the observed reduction in the asymptotic value of the VAO phase shift relative to the theoretical expectation of two, as well as transition scale $k_\star$ being further smoothed to smaller scales. 

\subsection{The Phase Shift in the 21-cm Power Spectrum}

\begin{figure*}
    \centering
    \includegraphics[width=\linewidth]{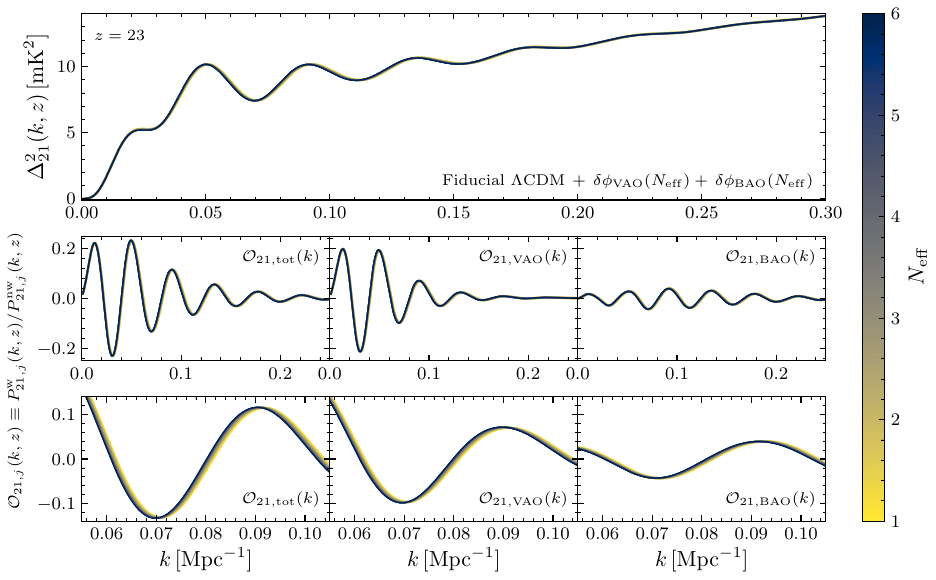}
    \caption{Illustration of the phase shift as it appears in the 21-cm power spectrum at the benchmark redshift $z=23$, corresponding to the midpoint of the LCE where $\overline{T}_{21}=0.5\times \overline{T}_{21,\rm min}$ (see Fig.~\ref{fig:6}).  {\bf Top Panel:} The reduced 21-cm power spectrum $\Delta_{21}^2(k,z) \equiv k^3 P_{21}(k,z)/(2\pi^2)$ computed using our fiducial $\Lambda$CDM model. The neutrino-induced phase shift has been imprinted into the source density and velocity spectra using the respective BAO/VAO best-fit templates for varying effective number of relativistic species $N_{\rm eff} \in [1,6]$.  This approach is a robust way to isolate the phase effect without the complications that would arise from self-consistently varying $N_{\rm eff}$. {\bf Middle and Bottom panels:} The corresponding full oscillatory 21-cm spectra $\mathcal{O}_{21,\rm tot}(k,z)$ ({\bf Left}) and their individual VAO. ({\bf Middle}), and BAO contributions ({\bf Right}), as defined in Eq.~\eqref{eq:O21_v0} and~\eqref{eq:O21}. The bottom row provides a zoomed-in view around a narrow k-range to further emphasize the phase shift. At this redshift, VAO contributions dominate at large scales while BAOs become more significant at small scales where Lyman-$\alpha$ photon propagation progressively dampens the velocity-induced oscillations.}
    \label{fig:4}
\end{figure*}

\noindent The phase shift in the 21-cm spectrum follows directly from our earlier decomposition of its acoustic oscillations into the BAO and VAO components, i.e. Eq.~\eqref{eq:O21}. The net phase shift $\phi_{21}$ is thus well approximated as a weighted sum of the individual contributions from BAOs and VAOs, respectively. In more detail:
\begin{align}
    \phi_{21}(k,z;N_{\rm eff})&\approx\mathcal{A}_{\rm BAO}(k,z) \phi_{\rm BAO}(k;N_{\rm eff}) \nonumber \\
    &+ \mathcal{A}_{\rm VAO}(k,z) \phi_{\rm VAO}(k;N_{\rm eff}),\label{eq:dphi21}
\end{align}
where $\mathcal{A}_{\rm BAO/VAO}(k,z)$ represent the relative amplitudes of each oscillatory component at mode $k$ and redshift $z$, with $\mathcal{A}_{\rm VAO}\approx\left(1- \mathcal{A}_{\rm BAO}\right)$, and the induced phases $\phi_{\rm BAO/VAO}$ are given by Eq.~\eqref{eq:dphi} for their respective best-fit templates $f_\phi$ as shown in Fig.~\ref{fig:3}. The validity of this weighted sum formulation can be demonstrated analytically. Following the standard qualitative treatment of the  acoustic oscillations in the primordial plasma, Eq.~\eqref{eq:phase0}, we can express each oscillatory component in the 21-cm spectrum as:
\begin{equation}
\mathcal{O}_{21,j}(k,z) = A_j(k,z)\sin(kr_s + \phi_j), \label{eq:O21_approx}
\end{equation}
where the index $j$ denotes either the individual components (BAO or VAO) or their combined signal, $A_j(k,z)$ is the corresponding $k-$dependent oscillation amplitude at redshift $z$, $r_s$ is the sound horizon, and $\phi_j$ captures the neutrino-induced phase shift. In the small-phase limit ($\phi_j \ll 1$), substituting Eq.~\eqref{eq:O21_approx} into the decomposition of Eq.~\eqref{eq:O21} directly yields the weighted sum in Eq.~\eqref{eq:dphi21}, with the relative amplitudes given by $\mathcal{A}_{\rm BAO}=A_{\rm BAO}/(A_{\rm VAO}+A_{\rm BAO})$ and $\mathcal{A}_{\rm VAO}=1 -\mathcal{A}_{\rm BAO}$. This analytical derivation confirms that the net phase shift is simply the amplitude-weighted average of the individual BAO and VAO phases, establishing the theoretical basis for our decomposition approach.

Given this linear superposition, we expect the neutrino-induced phase shift in the 21-cm spectrum to exhibit a distinctive mode and redshift dependence that reflects the relative evolution of VAO and BAO contributions -- a signature unique to this cosmological observable. Specifically, we expect small scales to be mostly BAO-dominated across all epochs due to the dominant contribution of density-sourced terms over velocity-sourced terms at high $k$~\citep{cruz25}, with further damping of VAOs from non-local photon propagation. In contrast, at large scales the phase shift should vary dramatically with redshift, tracking VAO signatures during the LCE when streaming velocities suppress star formation, then converging to the BAO behavior at late times as density-driven processes dominate, with the EoH introducing a temporary revival of VAO contributions driven by X-ray heating between these two regimes.

Figure~\ref{fig:4} illustrates the phase shift in the 21-cm spectrum and its separate BAO and VAO components at the benchmark redshift $z=23$, corresponding in our fiducial model to the halfway point through the LCE where $\overline{T}_{21}= 0.5 \times \overline{T}_{21,\rm min}$. At this redshift, both BAO and VAO contributions are significant at small and large scales, respectively. The 21-cm spectra are computed here using \texttt{zeus21}~\cite{munoz23,munoz23b,Cruz:2024fsv}, assuming our fiducial $\Lambda$CDM cosmology (Table~\ref{tab:parameters}) and manually imprinting the neutrino-induced phase shifts in the source density and velocity spectra according to their respective best-fit templates. This approach cleanly isolates the phase effects in the resulting 21-cm spectrum by avoiding the complications that would arise from self-consistently varying $N_{\rm eff}$ as a cosmological parameter. By applying this manual imprinting method across different redshifts and extracting the resulting mode shifts, we can directly derive the phase-shift template for the 21-cm spectrum.

We implement this procedure across the redshift range $z \in [10,30]$ by computing mode shifts $\delta k$ from the phase-shifted 21-cm oscillatory spectra and comparing against our fiducial model. To accurately propagate uncertainties from the source spectra, we sample the BAO and VAO phase shift parameters within their 95\% confidence intervals, propagating these errors into the final 21-cm phase shift measurements. We present the extracted phase-shift templates as smooth spline interpolations, across the full redshift range of interest in Fig.~\ref{fig:5} in the Appendix.

\begin{figure*}
    \centering
    \includegraphics[width=\linewidth]{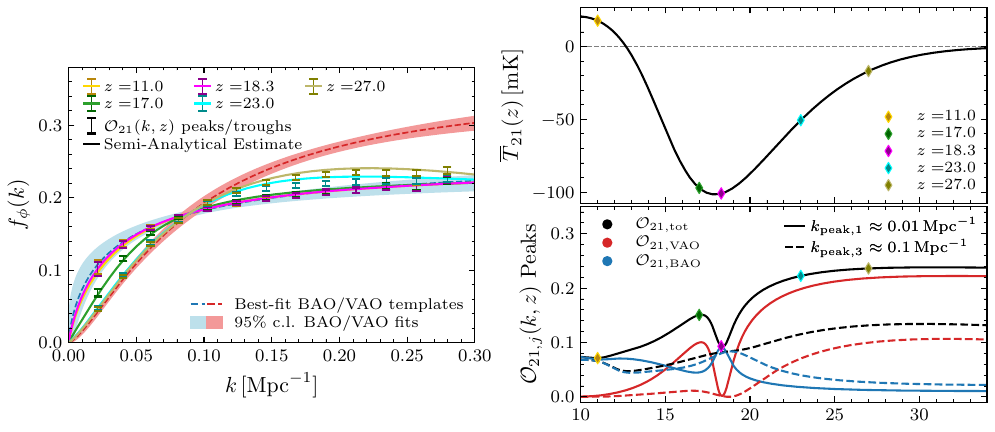}
    \caption{{\bf Left:} Phase shift-template in the 21-cm spectrum as a function of wavenumber $k$ at five critical redshifts that span the major astrophysical transitions during cosmic dawn. The numerical phase shifts for the 21-cm spectrum at the various redshifts, displayed in their respective colors as shown in the legend, were obtained by manually imprinting the neutrino-induced phase shifts in the source density and velocity spectra. The BAO and VAO templates used as inputs are shown as solid \textcolor{tabred}{red} and \textcolor{tabblue}{blue} lines respectively, with shaded regions indicating their $2\sigma$ confidence intervals. Error bars on the 21-cm phase shifts indicate the propagated uncertainties from sampling the BAO and VAO phase shift parameters within their 1$\sigma$ confidence intervals.  Solid lines show the semi-analytical predictions using our weighted sum formula, Eq.~\eqref{eq:dphi21}, which are in excellent agreement with the numerical results, corroborating our theoretical framework that the 21-cm phase shift is accurately modeled as a weighted superposition of the individual BAO and VAO phase templates. {\bf Top right:} Evolution of the average 21-cm brightness temperature across cosmic dawn for our fiducial model, with colored diamonds marking the five benchmark redshifts analyzed in the left panel. {\bf Bottom right:} Amplitude evolution of the first ($k \approx 0.01\,\rm{Mpc}^{-1}$) and third ($k\approx0.1\,\rm{Mpc}^{-1}$) acoustic peaks in the 21-cm oscillatory spectrum, Eq.~\eqref{eq:O21}, showing the total signal (black) and the individual BAO (\textcolor{tabblue}{blue}) and VAO (\textcolor{tabred}{red}) contributions. The relative BAO/VAO amplitudes at different scales directly determine the weights in our semi-analytical formula, illustrating the transition from VAO-dominated to BAO-dominated regimes across redshift and scale. }
    \label{fig:6}
\end{figure*}

For clarity, here in the main text, we specifically focus on the evolution of the 21-cm phase shift at five critical redshifts that span the major astrophysical transitions during cosmic dawn. In addition to the midpoint of the LCE at $z=23$, we examine four other characteristic redshifts: $z=27$, deep in the LCE where the relative VAO contribution remains maximized; $z=18.3$, roughly at the minimum of the average 21-cm temperature, where VAO amplitudes cancel due to the transition between the LCE and the EoH; $z=17$ corresponding approximately to the second VAO amplitude peak during the EoH; and finally $z=11$, when the IGM is fully heated and density-driven processes dominate. We display the numerically extracted phase shifts at these benchmark redshifts in Fig.~\ref{fig:6}. To complement these phase shift measurements, the right panels of the figure show the evolution of the average 21-cm temperature, as well as the amplitude of the first and third peak of the 21-cm oscillatory spectrum, including both the total signal and the individual BAO/VAO contributions.

Overall, these results confirm our theoretical expectations. At $z=27$ and $z=23$ (deep in and at the midpoint of the LCE, respectively), the phase shift closely follows the VAO template at large scales, as VAOs dominate the 21-cm signal through the suppression of star formation by the high streaming velocities. In contrast, at smaller scales ($k\gtrsim 0.15\,\rm{Mpc}^{-1}$), the phase transitions toward the BAO prediction as density-sourced terms from $P_m$ increasingly dominate over velocity-sourced contributions from $P_\eta$, as the latter decays with $k$ much faster than the former. Nevertheless, the phase remains elevated above the pure BAO expectation, particularly at $z=27$, where VAOs retain a non-negligible amplitude. For all other redshifts, small-scale modes consistently asymptote to the BAO phase as VAO contributions become increasingly suppressed by the longer photon mean free paths that develop with cosmic time, with the combined impact of Lyman-series and X-ray photons providing even stronger damping at lower redshifts ($z\lesssim 17$)~\cite{cruz24,cruz25}. This redshift and scale-dependent behavior is evident in the bottom-right panel of Fig.~\ref{fig:6}, where the BAO amplitude surpasses the VAO contribution already at their third peak ($k\approx 0.1\,\rm{Mpc}^{-1}$) for $z\lesssim 22$. In addition, the panel also shows the progressive damping of VAO contributions between the first and third peaks: at high redshift the VAO amplitude decreases by roughly a factor of 3 between these two scales, while at the VAO maximum during the EoH this reduction reaches already a factor of 10. This behavior reflects the combined impact of non-local photon propagation (from both X-rays and Lyman-$\alpha$) and in particular the increasing dominance of density-sourced terms from the matter spectrum over velocity-sourced contributions in shaping the small-scale 21-cm signal.

At both $z=18.3$ (the VAO cancellation redshift) and $z=11$ (fully heated IGM), the phase shift matches the BAO signature across all scales, as expected. The bottom-right panel of Fig.~\ref{fig:6} confirms negligible VAO contributions at these redshifts, though for different physical reasons: at $z=18.3$, near the 21-cm temperature minimum, the counteracting effects of relative velocities on Lyman-$\alpha$ and X-ray photon production precisely cancel, whereas at $z=11$, VAOs are fully damped by non-local photon propagation, leaving only density-driven BAO signatures. Finally, at the intermediate redshift $z=17$, corresponding to the second VAO maximum during the EoH, the phase partially shifts back towards the VAO template at large scales, though more moderately than during Lyman-$\alpha$ coupling as BAO contributions remain substantial  throughout. This scale dependence is explicitly demonstrated in the bottom-right panel of Fig.~\ref{fig:6}, where at $z=17$ the VAOs still dominate at the first acoustic peak ($k\approx 0.01\,\rm{Mpc}^{-1}$) despite non-negligible BAO contributions, yet transition to being strongly subdominant already by the third acoustic peak, reflecting their more efficient damping relative to BAOs during the EoH~\cite{cruz24,cruz25}. 

To demonstrate that our theoretical understanding accurately captures these complex dynamics, we also display in Fig.~\ref{fig:6} as solid lines the semi-analytical predictions using our weighted sum formula, Eq.~\eqref{eq:dphi21}. Here, we extract the amplitude ratios $\mathcal{A}_{\rm BAO}(k,z)$ by measuring the relative peak heights between the isolated BAO component and the full 21-cm oscillatory spectrum, then interpolating these values using a cubic spline and setting $\mathcal{A}_{\rm VAO} = 1 - \mathcal{A}_{\rm BAO}$. This approach maintains the essential physics since the amplitude fractions approximately sum to unity, while avoiding the challenging peak extraction from the heavily damped VAO component at small scales, which would introduce significant noise in our procedure.

As clearly depicted in the figure, the semi-analytical approximation closely reproduces the numerical phase shifts across all redshifts under consideration. In particular, it accurately captures the characteristic large-scale ($k\lesssim 0.1\,\rm{Mpc}^{-1}$) evolution from the VAO-dominated regime at $z=27$ through the VAO cancellation point at $z=18.3$, the partial return toward VAO signatures during the EoH at $z=17$, and the final convergence to pure BAO behavior at $z=11$. Minor deviations from our semi-analytical approach appear primarily at high redshifts and small scales, where both oscillatory components experience significant damping, yet the BAOs still constitute a substantial fraction ($\mathcal{A}_{\rm BAO} \sim 0.7-0.9$) of the total signal, making precise amplitude extraction more challenging in these transitional regimes.

While the specific redshifts of these characteristic epochs depend strongly on astrophysical parameters -- stellar formation efficiencies, feedback mechanisms, and X-ray heating rates can all enhance or suppress VAO contributions and modify their damping evolution -- the fundamental phase shift behavior remains robust. The neutrino-induced phase shift in the 21-cm power spectrum will consistently interpolate between the pure BAO and VAO signatures, with the relative amplitudes encoding the transition from streaming-velocity-dominated early times to density-driven late epochs. To illustrate both the robustness of this framework and the rich astrophysical information encoded in the phase shift evolution, we now examine a representative example of how varying key astrophysical parameters can affect this signature.

\subsubsection*{Astrophysical Modulation of the Phase Shift: The Role of Pop III Star Formation}
\label{sec:fstarIII}

\begin{figure}
    \centering
    \includegraphics[width=\linewidth]{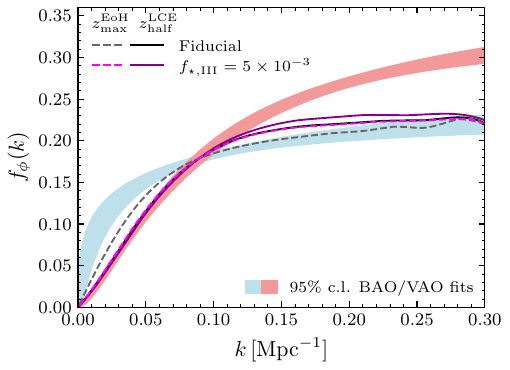}
    \caption{Comparison of the neutrino-induced phase shift in the 21-cm power spectrum between the fiducial model (\textcolor{tabgray}{gray}/ black) and the  enhanced $f_{\star,\rm III}$ scenario (\textcolor{magenta}{magenta}/ \textcolor{violet}{violet}), with $f_{\star,\rm III}=5\times f_{\star,\rm III}^{\rm fid}=5\times 10^{-3}$. The extracted phase shifts and smooth spline interpolations are shown at two representative epochs: the midpoint of the LCE, $z_{\rm half}^{\rm LCE}$ (darker curves), and and the VAO peak during the EoH, $z_{\rm max}^{\rm EoH}$ (lighter, dashed curves). For reference, we also display the $2\sigma$ confidence intervals of the BAO and VAO best-fit templates as shaded \textcolor{tabred}{red} and \textcolor{tabblue}{blue} regions respectively. At small scales ($k\gtrsim 0.15,\mathrm{Mpc}^{-1}$), both models approach the BAO prediction due to Lyman-$\alpha$ and X-ray damping, though the enhanced $f_{\star,\rm III}$ model remains systematically above the fiducial, reflecting its stronger VAO contributions. }
    \label{fig:7}
\end{figure}

\noindent As a concrete example of how astrophysical processes modulate the phase shift while preserving its fundamental character, we consider the impact of enhanced Pop III star formation efficiency. Specifically, motivated by recent hints from the James Webb Space Telescope~\citep{Venditti:2025mgi}, we examine a scenario where $f_{\star,\rm III}$ is increased by a factor of five relative to our fiducial model, i.e. we set $f_{\star,\rm III}=5\times 10^{-3}$. This enhancement, while still within current astrophysical uncertainties, significantly amplifies the contribution of mini-halos to early star formation. The increased Pop III activity produces a stronger Lyman-$\alpha$ radiation field that drives earlier and more efficient WF coupling, resulting in an earlier and deeper absorption trough in the global 21-cm signal, as shown in Fig.~\ref{fig:A2} in the Appendix. Crucially, this enhanced mini-halo contribution makes the 21-cm spectrum more sensitive to streaming-velocity modulation throughout both the LCE and the EoH, leading to markedly stronger VAO amplitudes compared to the fiducial case.

To quantify these effects, Figure~\ref{fig:7} displays the numerically extracted phase shifts and their corresponding smooth spline interpolations for both the fiducial and enhanced $f_{\star,\rm III}$ models at two critical redshifts.  Specifically, we examine (i) the midpoint of the LCE, $z_{\rm half}^{\rm LCE}$, occurring at $z=23$ and $z=26$ respectively for the fiducial and enhanced $f_{\star,\rm III}$ model; and (ii) the peak of the relative VAO contribution during the epoch of heating, $z_{\rm max}^{\rm EoH}$, occurring at $z=17$ in both scenarios. It is important to note that while this peak redshift remains unchanged, the relative VAO contribution during the EoH is significantly altered. The enhanced $f_{\star,\rm III}$ scenario maintains VAO dominance over a substantially extended period, roughly between $z\in[13,20]$, compared to the narrower window in the fiducial case where VAO contributions quickly diminish after their peak (see Fig.~\ref{fig:A2} in the Appendix).

Overall, the results illustrated in Fig.~\ref{fig:7} broadly confirm our theoretical expectations. At $z_{\rm half}^{\rm LCE}$, both the fiducial and enhanced $f_{\star,\rm III}$ models follow the VAO template closely at large scales, as VAO contributions remain dominant at these redshifts for both choices of the star formation efficiency. At smaller scales ($k \gtrsim 0.15\,\rm{Mpc}^{-1}$), while both models show the expected transition toward the BAO prediction as density-sourced terms from $P_m$ increasingly dominate over velocity-sourced contributions from $P_\eta$, the enhanced $f_{\star,\rm III}$ model systematically maintains  a higher phase shift. This elevation reflects the stronger velocity modulation imprinted by the increased mini-halo contribution, though the overall impact remains relatively modest during the LCE as both models ultimately converge toward similar small-scale behavior.

The differences become more pronounced at $z_{\rm max}^{\rm EoH}$, where the enhanced Pop III efficiency fundamentally alters the balance between BAO and VAO contributions. While small-scale modes still asymptote toward the BAO limit due to the combined damping effects of Lyman-series and X-ray photons, the enhanced $f_{\star \rm III}$ model maintains a phase shift significantly above the BAO expectation for all $k \lesssim 0.3\,\rm{Mpc}^{-1}$, with values comparable to those observed during the LCE. The large-scale behavior shows even more dramatic differences. In contrast to the fiducial case, where BAO contributions remain non-negligible throughout the EoH, the enhanced Pop III scenario produces complete VAO dominance at scales $k \lesssim 0.1\,\rm{Mpc}^{-1}$. 

These results highlight how the balance of BAO and VAO contributions -- and thus the detailed shape of the phase shift -- can significantly vary under plausible astrophysical scenarios, especially on large scales.  However, the inevitable damping of VAOs at small scales --  arising both from non-local photon propagation and especially the intrinsically smaller amplitude of the source velocity spectrum relative to the matter spectrum at high $k$ -- ensures that the phase shift in this regime remains anchored near the BAO prediction, making this asymptotic behavior a robust feature of the signal. Notably, though relatively close to the BAO limit, the phase shift remains elevated above the 95\% confidence interval of the BAO template even at intermediate scales (up to $k \sim 0.25\,\rm{Mpc}^{-1}$), particularly during the LCE, indicating that VAO contributions remain statistically significant in this regime and should not be neglected.

\section{Discussion \& Conclusions}\label{sec:5}

\noindent In this work, we have presented the first comprehensive analysis of the neutrino-induced phase shift in the 21-cm power spectrum during cosmic dawn, revealing a unique cosmological signature that emerges from the rich acoustic structure imprinted by both BAOs and VAOs, respectively sourced by matter density fluctuations and supersonic baryon-DM relative velocities. 

Central to understanding this signal is the derivation of the phase-shift template, Eq.~\eqref{eq:dphi}, in the VAO spectrum, which 
is one of our main results. As we show in Fig.~\ref{fig:3}, the phase in the VAOs significantly departs from that in the BAOs, which is fundamentally due to the four-point nature of the scalar power spectrum of the relative velocity $P_\eta(k)$, Eq.~\eqref{eq:Peta}, the source of the VAO features in the 21-cm signal. This is in contrast with the BAOs, which arise from standard two-point correlations of the matter and baryon density fields. Overall, we find that the VAO phase shift asymptotes to a value $\phi_\infty$ approximately $1.5$ times larger than the BAO phase shift and exhibits a smoother transition between large and small scales, resulting in a roughly tripled transition scale $k_\star$, see Eq.~\eqref{eq:fphi}. These unique features arise from the convolution of phase-shifted acoustic velocity modes in the four-point integral structure, additionally modified by the presence of a non-oscillatory component to the $\eta$ spectrum that dilutes the coherent phase addition.

Given these distinct BAO and VAO phase templates, we find that their combined effect produces a rich, evolving phase shift signature in the 21-cm power spectrum, reflecting the intricate interplay between BAO and VAO contributions as different astrophysical processes dominate at various epochs and scales across cosmic dawn, see Fig.~\ref{fig:4}. Our analysis demonstrates that this phase shift can be accurately modeled as a weighted superposition of the individual BAO and VAO phase templates, Eq.~\eqref{eq:dphi21}, with the relative amplitudes encoding the complex astrophysical evolution from streaming-velocity-dominated early times to density-driven late epochs. The sensitivity of the phase shift to the VAO/BAO amplitude ratio --ranging from nearly pure VAO signatures at large scales during Lyman-$\alpha$ coupling to complete BAO dominance at late times -- underscores the value of phase information in the 21-cm spectrum as a probe of both fundamental physics and astrophysical processes during cosmic dawn.

As a representative example of the impact of astrophysical parameters, we examined how an enhanced Pop III star formation efficiency modulates the phase shift signal, see Fig.~\ref{fig:7}. As expected, this enhancement significantly alters the balance of BAO and VAO contributions, particularly prolonging VAO dominance on large scales. Nevertheless, the inevitable damping of VAOs at small scales ensures that the phase shift consistently converges toward the BAO prediction in this regime, making this asymptotic behavior a robust feature of this signal.

Looking ahead, incorporating the phase information into 21-cm analyses offers a new, complementary probe alongside the well-established CMB and BAO measurements in large-scale structure surveys. The phase shift signal studied here should be viewed as one step toward a more complete understanding of the acoustic structure of the  21-cm signal. In particular, as emphasized in Ref.~\cite{Green:2020fjb}, the phase of the primordial acoustic oscillations is subject to additional contributions beyond the neutrino-induced effect. A systematic study of these effects in the context of the 21-cm line would clarify the extent to which different contributions can be disentangled, and how they may bias or enhance the interpretation of cosmological measurements. Despite these complexities, the prospects for detecting this phase shift signature are promising. Current and upcoming 21-cm experiments, including the operational Hydrogen Epoch of Reionization Array (HERA)~\citep{deboer17,hera23} and the soon-to-be-completed Square Kilometre Array (SKA)~\citep{SKA:2018ckk, SKA19}, will provide unprecedented sensitivity to the 21-cm power spectrum during cosmic dawn. Previous forecasts have demonstrated that VAOs themselves are detectable with these instruments~\cite{munoz19b, Sarkar:2022mdz}, suggesting that the phase shift signature we have characterized will also be within reach, though the phase detection specifically requires relatively large scales where measurements are challenging due to foregrounds~\cite{liu20}. 
While a quantitative sensitivity forecast is beyond the scope of this work, the framework developed here demonstrates that the neutrino-induced phase shift has a distinct and trackable imprint in the 21-cm power spectrum, providing a clear target for these ongoing and upcoming observational campaigns. In this way, the 21-cm phase shift can serve as valuable probe for both cosmology and astrophysics, through its sensitivity to the free-streaming nature of neutrinos and the astrophysical processes that determine the relative contributions of BAOs and VAOs throughout cosmic dawn.

\acknowledgments
We thank Benjamin Wallisch for valuable discussions. GM acknowledges support by the Continuing Fellowship of the Graduate School of the College of Natural Sciences at the University of Texas at Austin.  HAC was supported by the National Science Foundation Graduate Research Fellowship under Grant No.\ DGE2139757. 
 JBM was supported by NSF Grants AST-2307354 and AST-2408637, and by the NSF-Simons AI Institute for Cosmic Origins. 
 EDK acknowledges
joint support from the U.S.-Israel Bi-national Science
Foundation (BSF, grant No.\ 2022743) and the U.S.\ National Science Foundation (NSF, grant No.\ 2307354), as
well as support from the ISF-NSFC joint research program (grant No.\ 3156/23). EDK thanks the Weinberg Institute at University of Texas at Austin for hospitality during a sabbatical visit. This work was supported at JHU by NSF Grant No.\ 2412361, NASA ATP Grant No.\ 80NSSC24K1226, and the Templeton Foundation.
 We acknowledge the use of \texttt{CLASS}~\cite{Blas:2011rf},  \texttt{zeus21}~\cite{munoz23,munoz23b,Cruz:2024fsv}, and the Python packages \texttt{Matplotlib}~\cite{Hunter:2007mat}, \texttt{NumPy}~\cite{Harris:2020xlr} and~\texttt{SciPy}~\cite{Virtanen:2019joe}.

\appendix
\section{Extended Tests}
\begin{figure}
    \centering
    \includegraphics[width=\linewidth]{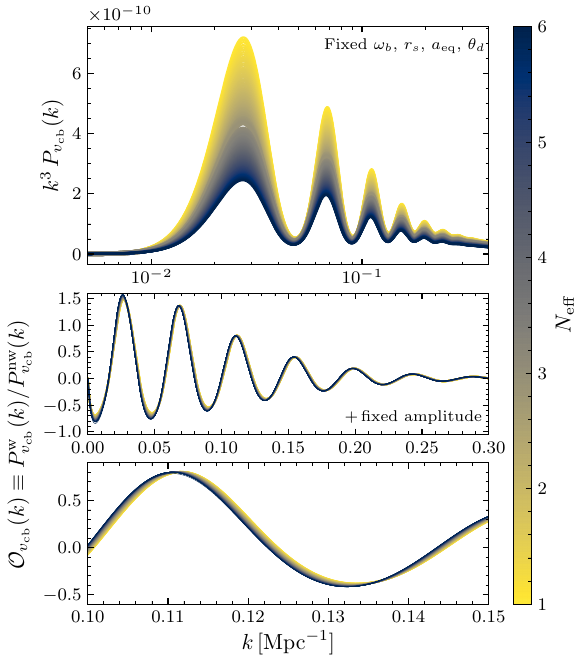}
    \caption{ Same as Fig.~\ref{fig:1} for the power spectrum of the baryon-DM relative velocity vector field, $P_{v_{\rm cb}}(k)$, Eq.~\eqref{eq:Pvcb}. As a two-point function of the underlying relative velocity perturbations, $P_{v_{\rm cb}}$ inherits as expected the standard BAO phase shift, as shown in Fig.~\ref{fig:3}. This is in contrast to the distinct response of the VAO features that arise from the scalar velocity spectrum $P_\eta$, Eq.~\eqref{eq:Peta}.}
    \label{fig:A1}
\end{figure}

\noindent This appendix presents three complementary figures that provide additional support for the main analysis of the neutrino-induced phase shift in the 21-cm power spectrum. In particular, we provide the equivalent of Fig.~\ref{fig:1} for the power spectrum of the baryon-DM relative velocity vector field, $P_{v_{\rm cb}}(k)$, illustrating explicitly the phase shift imprinted in this two-point function that serves as the building block for the scalar velocity spectrum $P_\eta$ which ultimately sources the VAOs imprinted in the 21-cm line. In addition, we display the complete redshift evolution of the 21-cm phase shift as smooth spline templates across cosmic dawn, corroborating our main conclusion that the phase shift continuously interpolates between the distinct BAO and VAO signatures as astrophysical conditions evolve. Finally, we provide a detailed representation of how the relative BAO/VAO amplitude evolution changes when varying astrophysical parameters, specifically the Pop III star formation efficiency, complementing the phase shift results already displayed in Fig.~\ref{fig:7} in the main text. 

\subsection{Phase shift in the velocity power spectrum.} 

\noindent Figure \ref{fig:A1} illustrates the extraction of the neutrino-induced phase shift from the baryon-DM relative velocity power spectrum $P_{v_{\rm cb}}(k)$, Eq.~\eqref{eq:Pvcb}, as a function of the effective number of relativistic species $N_{\rm eff}$. To isolate the phase shift, we follow the same procedure used for the matter and $\eta$ power spectra as shown in Fig.~\ref{fig:1}. The resulting phase shift, shown in purple in Fig.~\ref{fig:3}, matches the BAO template as expected, since the baryon velocity divergence $\theta_b$ is related to baryon density perturbations through a time derivative and thus experiences the same neutrino-induced phase shift prior to the drag epoch. This correspondence confirms that while $P_{v_{\rm cb}}$ inherits the standard BAO phase shift, the VAO features in the 21-cm spectrum will exhibit a distinct phase signature due to their origin in the scalar velocity power spectrum $P_\eta$, Eq.~\eqref{eq:Peta}, which instead represents effectively a four-point function of the baryon-DM relative velocity field, naturally leading to a modified phase response as derived in Section~\ref{sec:phase_vao}.

\begin{figure}
    \centering
    \includegraphics[width=\linewidth]{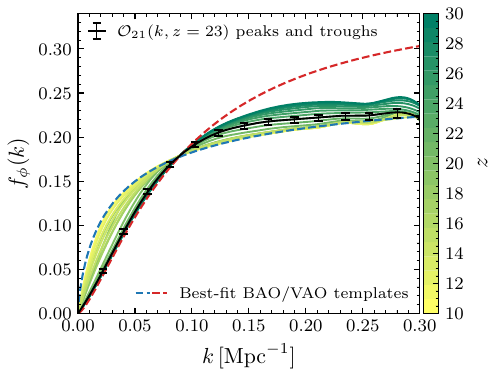}
    \caption{Extracted phase-shift templates for the 21-cm spectrum across cosmic dawn. The phase shifts are presented as smooth spline interpolations over the redshift range $z\in [10,30]$, displayed in a color gradient from yellow to green with increasing redshift. To illustrate the spline construction procedure, we show individual phase shift measurements with error bars at the peaks and troughs of the 21-cm oscillatory spectrum for the reference case at $z=23$ (corresponding to the midpoint of the Lyman-$\alpha$ coupling epoch).
    The dashed \textcolor{tabblue}{blue }and \textcolor{tabred}{red} lines show the best-fit BAO and VAO phase templates, respectively, between which the 21-cm phase shift interpolates with varying weights that reflect the complex interplay of BAO and VAO contributions throughout cosmic dawn.}
    \label{fig:5}
\end{figure}

\begin{figure*}[ht!]
    \centering
    \vspace{-.3cm}\includegraphics[width=\linewidth]{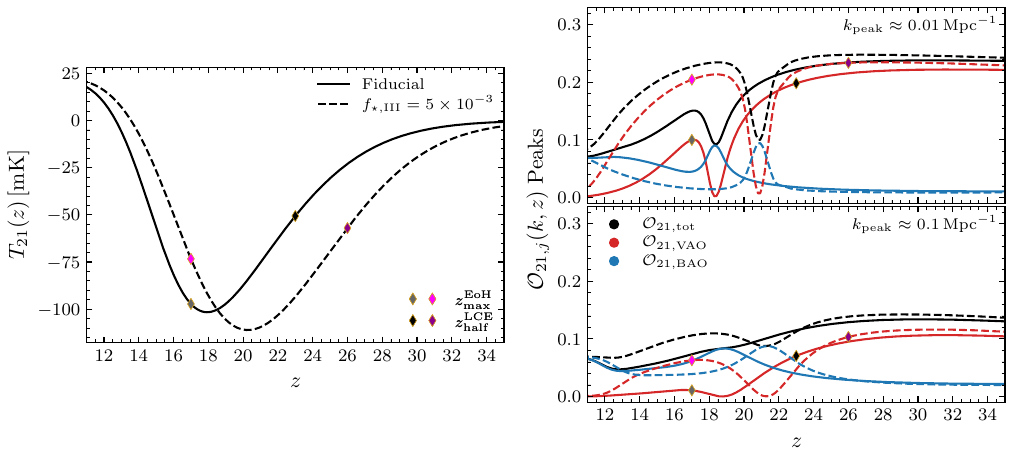}
    \caption{{\bf Left:} Evolution of the average 21-cm brightness temperature across cosmic dawn for our fiducial model (solid) and the enhanced $f_{\star,\mathrm{III}}$ case (dashed), with colored diamonds marking the midpoint of the LCE ($z_{\rm half}^{\rm LCE}$) and the peak of relative VAO contribution during the EoH ($z_{\rm max}^{\rm EoH}$) for each model. {\bf Right:} Amplitude evolution of the first ($k \approx 0.01\,\rm{Mpc}^{-1}$, {\bf top panel}) and third ($k\approx0.1\,\rm{Mpc}^{-1}$, {\bf bottom panel}) acoustic peaks in the 21-cm oscillatory spectrum, Eq.~\eqref{eq:O21}, showing the total signal (black) and the individual BAO (\textcolor{tabblue}{blue}) and VAO (\textcolor{tabred}{red}) contributions for both the fiducial (solid) and enhanced $f_{\star,\mathrm{III}}$ (dashed) models. The enhanced Pop III efficiency amplifies mini-halo contributions, leading to stronger VAO amplitudes and extended VAO dominance throughout the EoH, providing a concrete example of how astrophysical parameters modulate the relative BAO/VAO weights that determine the 21-cm phase shift.}
    \label{fig:A2}
\end{figure*}

\subsection{ Redshift evolution of the 21-cm phase-shift template.}
\noindent Figure~\ref{fig:5} presents the complete evolution of the 21-cm phase shift across cosmic dawn, spanning the full redshift range $z\in[10,30]$ analyzed in this work. The smooth spline interpolations, reveal how the phase shift continuously transitions between the BAO and VAO templates as astrophysical conditions evolve. This comprehensive view highlights the unique mode- and redshift-dependent signature that distinguishes the 21-cm phase shift from other cosmological probes. 

Looking toward future analyses of 21-cm observations, these spline templates could serve as the foundation for a generalized BAO/VAO fitting framework, where the amplitude of the phase shift becomes an additional free parameter -- analogous to approaches already implemented for galaxy survey BAO analyses in Ref.~\cite{Baumann:2017gkg,Baumann:2019keh} -- enabling direct constraints on $N_{\rm eff}$ from the acoustic features in the 21-cm spectrum.

\subsection{ Impact of astrophysical parameters} 
\noindent Figure~\ref{fig:A2} provides a concrete example of how varying astrophysical parameters modulates the phase shift while preserving its fundamental character, specifically examining how enhanced Pop III star formation efficiency alters the balance between BAO and VAO contributions as discussed in Section~\ref{sec:fstarIII}.

The left panel shows the evolution of the average 21-cm brightness temperature for both our fiducial model and the enhanced $f_{\star,\rm III}$ case, with diamonds marking two critical redshifts: the midpoint of the LCE ($z_{\rm half}^{\rm LCE}$) and the peak of relative VAO contribution during the EoH ($z_{\rm max}^{\rm EoH}$). The fivefold increase in $f_{\star,\rm III}$ drives earlier and stronger Lyman-$\alpha$ coupling through amplified mini-halo contributions, resulting in both a deeper absorption trough and extended VAO dominance during the epoch of heating.

The amplitude evolution at the first and third acoustic peaks, displayed in the right panels, clearly shows how the enhanced $f_{\star,\rm III}$ scenario maintains stronger VAO contributions across a broader redshift range, particularly for $z\in [13,20]$.  Note that $z_{\rm max}^{\rm EoH}$ is chosen to maximize VAO contributions across the broadest $k-$range for both models. While this choice is evident in the fiducial case, for the enhanced $f_{\star,\rm III}$, the first $k-$peak of the VAO contribution reaches its maximum at slightly earlier redshifts. Nevertheless, we can clearly see that for the third $k-$peak the VAO maximum is achieved precisely at the selected $z_{\rm max}^{\rm EoH}$, confirming that this redshift optimizes VAO dominance across all relevant scales. Despite these enhanced VAO contributions, the damping remains significant, especially during the EoH, with VAO amplitudes decreasing by nearly an order of magnitude between the first and third peaks, similarly to the fiducial model.
\vspace{-.5cm}
\bibliographystyle{apsrev4-1}
\bibliography{bibl}

\begin{thebibliography}{84}%
\makeatletter
\providecommand \@ifxundefined [1]{%
 \@ifx{#1\undefined}
}%
\providecommand \@ifnum [1]{%
 \ifnum #1\expandafter \@firstoftwo
 \else \expandafter \@secondoftwo
 \fi
}%
\providecommand \@ifx [1]{%
 \ifx #1\expandafter \@firstoftwo
 \else \expandafter \@secondoftwo
 \fi
}%
\providecommand \natexlab [1]{#1}%
\providecommand \enquote  [1]{``#1''}%
\providecommand \bibnamefont  [1]{#1}%
\providecommand \bibfnamefont [1]{#1}%
\providecommand \citenamefont [1]{#1}%
\providecommand \href@noop [0]{\@secondoftwo}%
\providecommand \href [0]{\begingroup \@sanitize@url \@href}%
\providecommand \@href[1]{\@@startlink{#1}\@@href}%
\providecommand \@@href[1]{\endgroup#1\@@endlink}%
\providecommand \@sanitize@url [0]{\catcode `\\12\catcode `\$12\catcode `\&12\catcode `\#12\catcode `\^12\catcode `\_12\catcode `\%12\relax}%
\providecommand \@@startlink[1]{}%
\providecommand \@@endlink[0]{}%
\providecommand \url  [0]{\begingroup\@sanitize@url \@url }%
\providecommand \@url [1]{\endgroup\@href {#1}{\urlprefix }}%
\providecommand \urlprefix  [0]{URL }%
\providecommand \Eprint [0]{\href }%
\providecommand \doibase [0]{http://dx.doi.org/}%
\providecommand \selectlanguage [0]{\@gobble}%
\providecommand \bibinfo  [0]{\@secondoftwo}%
\providecommand \bibfield  [0]{\@secondoftwo}%
\providecommand \translation [1]{[#1]}%
\providecommand \BibitemOpen [0]{}%
\providecommand \bibitemStop [0]{}%
\providecommand \bibitemNoStop [0]{.\EOS\space}%
\providecommand \EOS [0]{\spacefactor3000\relax}%
\providecommand \BibitemShut  [1]{\csname bibitem#1\endcsname}%
\let\auto@bib@innerbib\@empty
\bibitem [{\citenamefont {Annis}\ \emph {et~al.}(2022)\citenamefont {Annis}, \citenamefont {Newman},\ and\ \citenamefont {Slosar}}]{Annis:2022xgg}%
  \BibitemOpen
  \bibfield  {author} {\bibinfo {author} {\bibfnamefont {J.}~\bibnamefont {Annis}}, \bibinfo {author} {\bibfnamefont {J.}~\bibnamefont {Newman}}, \ and\ \bibinfo {author} {\bibfnamefont {A.}~\bibnamefont {Slosar}},\ }\href@noop {} {\  (\bibinfo {year} {2022})},\ \Eprint {http://arxiv.org/abs/2209.08049} {arXiv:2209.08049 [astro-ph.CO]} \BibitemShut {NoStop}%
\bibitem [{\citenamefont {Chang}\ \emph {et~al.}(2022)\citenamefont {Chang} \emph {et~al.}}]{Chang:2022lrw}%
  \BibitemOpen
  \bibfield  {author} {\bibinfo {author} {\bibfnamefont {C.}~\bibnamefont {Chang}} \emph {et~al.},\ }\href@noop {} {\  (\bibinfo {year} {2022})},\ \Eprint {http://arxiv.org/abs/2209.08265} {arXiv:2209.08265 [hep-ex]} \BibitemShut {NoStop}%
\bibitem [{\citenamefont {{Visbal}}\ \emph {et~al.}(2012)\citenamefont {{Visbal}}, \citenamefont {{Barkana}}, \citenamefont {{Fialkov}}, \citenamefont {{Tseliakhovich}},\ and\ \citenamefont {{Hirata}}}]{visbal12}%
  \BibitemOpen
  \bibfield  {author} {\bibinfo {author} {\bibfnamefont {E.}~\bibnamefont {{Visbal}}}, \bibinfo {author} {\bibfnamefont {R.}~\bibnamefont {{Barkana}}}, \bibinfo {author} {\bibfnamefont {A.}~\bibnamefont {{Fialkov}}}, \bibinfo {author} {\bibfnamefont {D.}~\bibnamefont {{Tseliakhovich}}}, \ and\ \bibinfo {author} {\bibfnamefont {C.~M.}\ \bibnamefont {{Hirata}}},\ }\href {\doibase 10.1038/nature11177} {\bibfield  {journal} {\bibinfo  {journal} {\nat}\ }\textbf {\bibinfo {volume} {487}},\ \bibinfo {pages} {70} (\bibinfo {year} {2012})},\ \Eprint {http://arxiv.org/abs/1201.1005} {arXiv:1201.1005 [astro-ph.CO]} \BibitemShut {NoStop}%
\bibitem [{\citenamefont {{McQuinn}}\ and\ \citenamefont {{O'Leary}}(2012)}]{mcquinn12}%
  \BibitemOpen
  \bibfield  {author} {\bibinfo {author} {\bibfnamefont {M.}~\bibnamefont {{McQuinn}}}\ and\ \bibinfo {author} {\bibfnamefont {R.~M.}\ \bibnamefont {{O'Leary}}},\ }\href {\doibase 10.1088/0004-637X/760/1/3} {\bibfield  {journal} {\bibinfo  {journal} {\apj}\ }\textbf {\bibinfo {volume} {760}},\ \bibinfo {eid} {3} (\bibinfo {year} {2012})},\ \Eprint {http://arxiv.org/abs/1204.1345} {arXiv:1204.1345 [astro-ph.CO]} \BibitemShut {NoStop}%
\bibitem [{\citenamefont {{Fialkov}}\ \emph {et~al.}(2012)\citenamefont {{Fialkov}}, \citenamefont {{Barkana}}, \citenamefont {{Tseliakhovich}},\ and\ \citenamefont {{Hirata}}}]{fialkov12}%
  \BibitemOpen
  \bibfield  {author} {\bibinfo {author} {\bibfnamefont {A.}~\bibnamefont {{Fialkov}}}, \bibinfo {author} {\bibfnamefont {R.}~\bibnamefont {{Barkana}}}, \bibinfo {author} {\bibfnamefont {D.}~\bibnamefont {{Tseliakhovich}}}, \ and\ \bibinfo {author} {\bibfnamefont {C.~M.}\ \bibnamefont {{Hirata}}},\ }\href {\doibase 10.1111/j.1365-2966.2012.21318.x} {\bibfield  {journal} {\bibinfo  {journal} {\mnras}\ }\textbf {\bibinfo {volume} {424}},\ \bibinfo {pages} {1335} (\bibinfo {year} {2012})},\ \Eprint {http://arxiv.org/abs/1110.2111} {arXiv:1110.2111 [astro-ph.CO]} \BibitemShut {NoStop}%
\bibitem [{\citenamefont {Fialkov}\ \emph {et~al.}(2014)\citenamefont {Fialkov}, \citenamefont {Barkana}, \citenamefont {Pinhas},\ and\ \citenamefont {Visbal}}]{Fialkov:2013uwm}%
  \BibitemOpen
  \bibfield  {author} {\bibinfo {author} {\bibfnamefont {A.}~\bibnamefont {Fialkov}}, \bibinfo {author} {\bibfnamefont {R.}~\bibnamefont {Barkana}}, \bibinfo {author} {\bibfnamefont {A.}~\bibnamefont {Pinhas}}, \ and\ \bibinfo {author} {\bibfnamefont {E.}~\bibnamefont {Visbal}},\ }\href {\doibase 10.1093/mnrasl/slt135} {\bibfield  {journal} {\bibinfo  {journal} {Mon. Not. Roy. Astron. Soc.}\ }\textbf {\bibinfo {volume} {437}},\ \bibinfo {pages} {36} (\bibinfo {year} {2014})},\ \Eprint {http://arxiv.org/abs/1306.2354} {arXiv:1306.2354 [astro-ph.CO]} \BibitemShut {NoStop}%
\bibitem [{\citenamefont {{Mu{\~n}oz}}(2019{\natexlab{a}})}]{munoz19}%
  \BibitemOpen
  \bibfield  {author} {\bibinfo {author} {\bibfnamefont {J.~B.}\ \bibnamefont {{Mu{\~n}oz}}},\ }\href {\doibase 10.1103/PhysRevD.100.063538} {\bibfield  {journal} {\bibinfo  {journal} {\prd}\ }\textbf {\bibinfo {volume} {100}},\ \bibinfo {eid} {063538} (\bibinfo {year} {2019}{\natexlab{a}})},\ \Eprint {http://arxiv.org/abs/1904.07881} {arXiv:1904.07881 [astro-ph.CO]} \BibitemShut {NoStop}%
\bibitem [{\citenamefont {{Mu{\~n}oz}}(2019{\natexlab{b}})}]{munoz19b}%
  \BibitemOpen
  \bibfield  {author} {\bibinfo {author} {\bibfnamefont {J.~B.}\ \bibnamefont {{Mu{\~n}oz}}},\ }\href {\doibase 10.1103/PhysRevLett.123.131301} {\bibfield  {journal} {\bibinfo  {journal} {\prl}\ }\textbf {\bibinfo {volume} {123}},\ \bibinfo {eid} {131301} (\bibinfo {year} {2019}{\natexlab{b}})},\ \Eprint {http://arxiv.org/abs/1904.07868} {arXiv:1904.07868 [astro-ph.CO]} \BibitemShut {NoStop}%
\bibitem [{\citenamefont {Sarkar}\ and\ \citenamefont {Kovetz}(2023)}]{Sarkar:2022mdz}%
  \BibitemOpen
  \bibfield  {author} {\bibinfo {author} {\bibfnamefont {D.}~\bibnamefont {Sarkar}}\ and\ \bibinfo {author} {\bibfnamefont {E.~D.}\ \bibnamefont {Kovetz}},\ }\href {\doibase 10.1103/PhysRevD.107.023524} {\bibfield  {journal} {\bibinfo  {journal} {Phys. Rev. D}\ }\textbf {\bibinfo {volume} {107}},\ \bibinfo {pages} {023524} (\bibinfo {year} {2023})},\ \Eprint {http://arxiv.org/abs/2210.16853} {arXiv:2210.16853 [astro-ph.CO]} \BibitemShut {NoStop}%
\bibitem [{\citenamefont {Cruz}\ \emph {et~al.}()\citenamefont {Cruz}, \citenamefont {Montefalcone}, \citenamefont {Munoz}, \citenamefont {Kovetz},\ and\ \citenamefont {Kamionkowski}}]{cruz25}%
  \BibitemOpen
  \bibfield  {author} {\bibinfo {author} {\bibfnamefont {H.~A.~G.}\ \bibnamefont {Cruz}}, \bibinfo {author} {\bibfnamefont {G.}~\bibnamefont {Montefalcone}}, \bibinfo {author} {\bibfnamefont {J.~B.}\ \bibnamefont {Munoz}}, \bibinfo {author} {\bibfnamefont {E.~D.}\ \bibnamefont {Kovetz}}, \ and\ \bibinfo {author} {\bibfnamefont {M.}~\bibnamefont {Kamionkowski}},\ }\href@noop {} {\enquote {\bibinfo {title} {{\it in preparation}},}\ }\BibitemShut {NoStop}%
\bibitem [{\citenamefont {{Mu{\~n}oz}}\ \emph {et~al.}(2020)\citenamefont {{Mu{\~n}oz}}, \citenamefont {{Dvorkin}},\ and\ \citenamefont {{Cyr-Racine}}}]{munoz20}%
  \BibitemOpen
  \bibfield  {author} {\bibinfo {author} {\bibfnamefont {J.~B.}\ \bibnamefont {{Mu{\~n}oz}}}, \bibinfo {author} {\bibfnamefont {C.}~\bibnamefont {{Dvorkin}}}, \ and\ \bibinfo {author} {\bibfnamefont {F.-Y.}\ \bibnamefont {{Cyr-Racine}}},\ }\href {\doibase 10.1103/PhysRevD.101.063526} {\bibfield  {journal} {\bibinfo  {journal} {\prd}\ }\textbf {\bibinfo {volume} {101}},\ \bibinfo {eid} {063526} (\bibinfo {year} {2020})},\ \Eprint {http://arxiv.org/abs/1911.11144} {arXiv:1911.11144 [astro-ph.CO]} \BibitemShut {NoStop}%
\bibitem [{\citenamefont {Libanore}\ \emph {et~al.}(2024)\citenamefont {Libanore}, \citenamefont {Flitter}, \citenamefont {Kovetz}, \citenamefont {Li},\ and\ \citenamefont {Dekel}}]{Libanore:2023oxf}%
  \BibitemOpen
  \bibfield  {author} {\bibinfo {author} {\bibfnamefont {S.}~\bibnamefont {Libanore}}, \bibinfo {author} {\bibfnamefont {J.}~\bibnamefont {Flitter}}, \bibinfo {author} {\bibfnamefont {E.~D.}\ \bibnamefont {Kovetz}}, \bibinfo {author} {\bibfnamefont {Z.}~\bibnamefont {Li}}, \ and\ \bibinfo {author} {\bibfnamefont {A.}~\bibnamefont {Dekel}},\ }\href {\doibase 10.1093/mnras/stae1485} {\bibfield  {journal} {\bibinfo  {journal} {Mon. Not. Roy. Astron. Soc.}\ }\textbf {\bibinfo {volume} {532}},\ \bibinfo {pages} {149} (\bibinfo {year} {2024})},\ \Eprint {http://arxiv.org/abs/2310.03021} {arXiv:2310.03021 [astro-ph.CO]} \BibitemShut {NoStop}%
\bibitem [{\citenamefont {Zhang}\ \emph {et~al.}(2024)\citenamefont {Zhang}, \citenamefont {Lin}, \citenamefont {Zhang}, \citenamefont {Yue}, \citenamefont {Gong}, \citenamefont {Xu},\ and\ \citenamefont {Chen}}]{Zhang:2024pwv}%
  \BibitemOpen
  \bibfield  {author} {\bibinfo {author} {\bibfnamefont {X.}~\bibnamefont {Zhang}}, \bibinfo {author} {\bibfnamefont {H.}~\bibnamefont {Lin}}, \bibinfo {author} {\bibfnamefont {M.}~\bibnamefont {Zhang}}, \bibinfo {author} {\bibfnamefont {B.}~\bibnamefont {Yue}}, \bibinfo {author} {\bibfnamefont {Y.}~\bibnamefont {Gong}}, \bibinfo {author} {\bibfnamefont {Y.}~\bibnamefont {Xu}}, \ and\ \bibinfo {author} {\bibfnamefont {X.}~\bibnamefont {Chen}},\ }\href {\doibase 10.3847/1538-4357/ad235b} {\bibfield  {journal} {\bibinfo  {journal} {Astrophys. J.}\ }\textbf {\bibinfo {volume} {964}},\ \bibinfo {pages} {62} (\bibinfo {year} {2024})},\ \Eprint {http://arxiv.org/abs/2401.14234} {arXiv:2401.14234 [astro-ph.CO]} \BibitemShut {NoStop}%
\bibitem [{\citenamefont {{Verwohlt}}\ \emph {et~al.}(2024)\citenamefont {{Verwohlt}}, \citenamefont {{Mason}}, \citenamefont {{Mu{\~n}oz}}, \citenamefont {{Cyr-Racine}}, \citenamefont {{Vogelsberger}},\ and\ \citenamefont {{Zavala}}}]{verwohlt2024}%
  \BibitemOpen
  \bibfield  {author} {\bibinfo {author} {\bibfnamefont {J.}~\bibnamefont {{Verwohlt}}}, \bibinfo {author} {\bibfnamefont {C.~A.}\ \bibnamefont {{Mason}}}, \bibinfo {author} {\bibfnamefont {J.~B.}\ \bibnamefont {{Mu{\~n}oz}}}, \bibinfo {author} {\bibfnamefont {F.-Y.}\ \bibnamefont {{Cyr-Racine}}}, \bibinfo {author} {\bibfnamefont {M.}~\bibnamefont {{Vogelsberger}}}, \ and\ \bibinfo {author} {\bibfnamefont {J.}~\bibnamefont {{Zavala}}},\ }\href {\doibase 10.1103/PhysRevD.110.103533} {\bibfield  {journal} {\bibinfo  {journal} {\prd}\ }\textbf {\bibinfo {volume} {110}},\ \bibinfo {eid} {103533} (\bibinfo {year} {2024})},\ \Eprint {http://arxiv.org/abs/2404.17640} {arXiv:2404.17640 [astro-ph.CO]} \BibitemShut {NoStop}%
\bibitem [{\citenamefont {Hotinli}\ \emph {et~al.}(2022)\citenamefont {Hotinli}, \citenamefont {Marsh},\ and\ \citenamefont {Kamionkowski}}]{Hotinli:2021vxg}%
  \BibitemOpen
  \bibfield  {author} {\bibinfo {author} {\bibfnamefont {S.~C.}\ \bibnamefont {Hotinli}}, \bibinfo {author} {\bibfnamefont {D.~J.~E.}\ \bibnamefont {Marsh}}, \ and\ \bibinfo {author} {\bibfnamefont {M.}~\bibnamefont {Kamionkowski}},\ }\href {\doibase 10.1103/PhysRevD.106.043529} {\bibfield  {journal} {\bibinfo  {journal} {Phys. Rev. D}\ }\textbf {\bibinfo {volume} {106}},\ \bibinfo {pages} {043529} (\bibinfo {year} {2022})},\ \Eprint {http://arxiv.org/abs/2112.06943} {arXiv:2112.06943 [astro-ph.CO]} \BibitemShut {NoStop}%
\bibitem [{\citenamefont {Hotinli}\ \emph {et~al.}(2021)\citenamefont {Hotinli}, \citenamefont {Binnie}, \citenamefont {Mu\~noz}, \citenamefont {Dinda},\ and\ \citenamefont {Kamionkowski}}]{Hotinli:2021xln}%
  \BibitemOpen
  \bibfield  {author} {\bibinfo {author} {\bibfnamefont {S.~C.}\ \bibnamefont {Hotinli}}, \bibinfo {author} {\bibfnamefont {T.}~\bibnamefont {Binnie}}, \bibinfo {author} {\bibfnamefont {J.~B.}\ \bibnamefont {Mu\~noz}}, \bibinfo {author} {\bibfnamefont {B.~R.}\ \bibnamefont {Dinda}}, \ and\ \bibinfo {author} {\bibfnamefont {M.}~\bibnamefont {Kamionkowski}},\ }\href {\doibase 10.1103/PhysRevD.104.063536} {\bibfield  {journal} {\bibinfo  {journal} {Phys. Rev. D}\ }\textbf {\bibinfo {volume} {104}},\ \bibinfo {pages} {063536} (\bibinfo {year} {2021})},\ \Eprint {http://arxiv.org/abs/2106.11979} {arXiv:2106.11979 [astro-ph.CO]} \BibitemShut {NoStop}%
\bibitem [{\citenamefont {{Cruz}}\ \emph {et~al.}(2024)\citenamefont {{Cruz}}, \citenamefont {{Adi}}, \citenamefont {{Flitter}}, \citenamefont {{Kamionkowski}},\ and\ \citenamefont {{Kovetz}}}]{cruz24}%
  \BibitemOpen
  \bibfield  {author} {\bibinfo {author} {\bibfnamefont {H.~A.~G.}\ \bibnamefont {{Cruz}}}, \bibinfo {author} {\bibfnamefont {T.}~\bibnamefont {{Adi}}}, \bibinfo {author} {\bibfnamefont {J.}~\bibnamefont {{Flitter}}}, \bibinfo {author} {\bibfnamefont {M.}~\bibnamefont {{Kamionkowski}}}, \ and\ \bibinfo {author} {\bibfnamefont {E.~D.}\ \bibnamefont {{Kovetz}}},\ }\href {\doibase 10.1103/PhysRevD.109.023518} {\bibfield  {journal} {\bibinfo  {journal} {\prd}\ }\textbf {\bibinfo {volume} {109}},\ \bibinfo {eid} {023518} (\bibinfo {year} {2024})},\ \Eprint {http://arxiv.org/abs/2308.04483} {arXiv:2308.04483 [astro-ph.CO]} \BibitemShut {NoStop}%
\bibitem [{\citenamefont {{Schauer}}\ \emph {et~al.}(2019)\citenamefont {{Schauer}}, \citenamefont {{Glover}}, \citenamefont {{Klessen}},\ and\ \citenamefont {{Ceverino}}}]{schauer19}%
  \BibitemOpen
  \bibfield  {author} {\bibinfo {author} {\bibfnamefont {A.~T.~P.}\ \bibnamefont {{Schauer}}}, \bibinfo {author} {\bibfnamefont {S.~C.~O.}\ \bibnamefont {{Glover}}}, \bibinfo {author} {\bibfnamefont {R.~S.}\ \bibnamefont {{Klessen}}}, \ and\ \bibinfo {author} {\bibfnamefont {D.}~\bibnamefont {{Ceverino}}},\ }\href {\doibase 10.1093/mnras/stz013} {\bibfield  {journal} {\bibinfo  {journal} {\mnras}\ }\textbf {\bibinfo {volume} {484}},\ \bibinfo {pages} {3510} (\bibinfo {year} {2019})},\ \Eprint {http://arxiv.org/abs/1811.12920} {arXiv:1811.12920 [astro-ph.GA]} \BibitemShut {NoStop}%
\bibitem [{\citenamefont {{Kulkarni}}\ \emph {et~al.}(2021)\citenamefont {{Kulkarni}}, \citenamefont {{Visbal}},\ and\ \citenamefont {{Bryan}}}]{kulkarni21}%
  \BibitemOpen
  \bibfield  {author} {\bibinfo {author} {\bibfnamefont {M.}~\bibnamefont {{Kulkarni}}}, \bibinfo {author} {\bibfnamefont {E.}~\bibnamefont {{Visbal}}}, \ and\ \bibinfo {author} {\bibfnamefont {G.~L.}\ \bibnamefont {{Bryan}}},\ }\href {\doibase 10.3847/1538-4357/ac08a3} {\bibfield  {journal} {\bibinfo  {journal} {\apj}\ }\textbf {\bibinfo {volume} {917}},\ \bibinfo {eid} {40} (\bibinfo {year} {2021})},\ \Eprint {http://arxiv.org/abs/2010.04169} {arXiv:2010.04169 [astro-ph.GA]} \BibitemShut {NoStop}%
\bibitem [{\citenamefont {Bauer}\ and\ \citenamefont {Shergold}(2023)}]{Bauer:2022lri}%
  \BibitemOpen
  \bibfield  {author} {\bibinfo {author} {\bibfnamefont {M.}~\bibnamefont {Bauer}}\ and\ \bibinfo {author} {\bibfnamefont {J.}~\bibnamefont {Shergold}},\ }\href {\doibase 10.1088/1475-7516/2023/01/003} {\bibfield  {journal} {\bibinfo  {journal} {JCAP}\ }\textbf {\bibinfo {volume} {01}},\ \bibinfo {pages} {003} (\bibinfo {year} {2023})},\ \Eprint {http://arxiv.org/abs/2207.12413} {arXiv:2207.12413 [hep-ph]} \BibitemShut {NoStop}%
\bibitem [{\citenamefont {{M. Betti \textit{et al.} (PTOLEMY Collaboration)}}(2019)}]{PTOLEMY:2019hkd}%
  \BibitemOpen
  \bibfield  {author} {\bibinfo {author} {\bibnamefont {{M. Betti \textit{et al.} (PTOLEMY Collaboration)}}},\ }\href {\doibase 10.1088/1475-7516/2019/07/047} {\bibfield  {journal} {\bibinfo  {journal} {JCAP}\ }\textbf {\bibinfo {volume} {07}},\ \bibinfo {pages} {047} (\bibinfo {year} {2019})},\ \Eprint {http://arxiv.org/abs/1902.05508} {arXiv:1902.05508 [astro-ph.CO]} \BibitemShut {NoStop}%
\bibitem [{\citenamefont {Bashinsky}\ and\ \citenamefont {Seljak}(2004)}]{Bashinsky:2003tk}%
  \BibitemOpen
  \bibfield  {author} {\bibinfo {author} {\bibfnamefont {S.}~\bibnamefont {Bashinsky}}\ and\ \bibinfo {author} {\bibfnamefont {U.}~\bibnamefont {Seljak}},\ }\href {\doibase 10.1103/PhysRevD.69.083002} {\bibfield  {journal} {\bibinfo  {journal} {Phys. Rev. D}\ }\textbf {\bibinfo {volume} {69}},\ \bibinfo {pages} {083002} (\bibinfo {year} {2004})},\ \Eprint {http://arxiv.org/abs/astro-ph/0310198} {arXiv:astro-ph/0310198 [astro-ph]} \BibitemShut {NoStop}%
\bibitem [{\citenamefont {Baumann}\ \emph {et~al.}(2016)\citenamefont {Baumann}, \citenamefont {Green}, \citenamefont {Meyers},\ and\ \citenamefont {Wallisch}}]{Baumann:2015rya}%
  \BibitemOpen
  \bibfield  {author} {\bibinfo {author} {\bibfnamefont {D.}~\bibnamefont {Baumann}}, \bibinfo {author} {\bibfnamefont {D.}~\bibnamefont {Green}}, \bibinfo {author} {\bibfnamefont {J.}~\bibnamefont {Meyers}}, \ and\ \bibinfo {author} {\bibfnamefont {B.}~\bibnamefont {Wallisch}},\ }\href {\doibase 10.1088/1475-7516/2016/01/007} {\bibfield  {journal} {\bibinfo  {journal} {JCAP}\ }\textbf {\bibinfo {volume} {01}},\ \bibinfo {pages} {007} (\bibinfo {year} {2016})},\ \Eprint {http://arxiv.org/abs/1508.06342} {arXiv:1508.06342 [astro-ph.CO]} \BibitemShut {NoStop}%
\bibitem [{\citenamefont {Follin}\ \emph {et~al.}(2015)\citenamefont {Follin}, \citenamefont {Knox}, \citenamefont {Millea},\ and\ \citenamefont {Pan}}]{Follin:2015hya}%
  \BibitemOpen
  \bibfield  {author} {\bibinfo {author} {\bibfnamefont {B.}~\bibnamefont {Follin}}, \bibinfo {author} {\bibfnamefont {L.}~\bibnamefont {Knox}}, \bibinfo {author} {\bibfnamefont {M.}~\bibnamefont {Millea}}, \ and\ \bibinfo {author} {\bibfnamefont {Z.}~\bibnamefont {Pan}},\ }\href {\doibase 10.1103/PhysRevLett.115.091301} {\bibfield  {journal} {\bibinfo  {journal} {Phys. Rev. Lett.}\ }\textbf {\bibinfo {volume} {115}},\ \bibinfo {pages} {091301} (\bibinfo {year} {2015})},\ \Eprint {http://arxiv.org/abs/1503.07863} {arXiv:1503.07863 [astro-ph.CO]} \BibitemShut {NoStop}%
\bibitem [{\citenamefont {Montefalcone}\ \emph {et~al.}(2025)\citenamefont {Montefalcone}, \citenamefont {Wallisch},\ and\ \citenamefont {Freese}}]{Montefalcone:2025unv}%
  \BibitemOpen
  \bibfield  {author} {\bibinfo {author} {\bibfnamefont {G.}~\bibnamefont {Montefalcone}}, \bibinfo {author} {\bibfnamefont {B.}~\bibnamefont {Wallisch}}, \ and\ \bibinfo {author} {\bibfnamefont {K.}~\bibnamefont {Freese}},\ }\href@noop {} {\enquote {\bibinfo {title} {{Free-Streaming Neutrinos and Their Phase Shift in Current and Future CMB Power Spectra}},}\ } (\bibinfo {year} {2025}),\ \Eprint {http://arxiv.org/abs/2501.13788} {arXiv:2501.13788 [astro-ph.CO]} \BibitemShut {NoStop}%
\bibitem [{\citenamefont {Montefalcone}\ \emph {et~al.}()\citenamefont {Montefalcone}, \citenamefont {Ghosh}, \citenamefont {Boddy}, \citenamefont {Wei Ren~Ho},\ and\ \citenamefont {Tsai}}]{montefalcone25}%
  \BibitemOpen
  \bibfield  {author} {\bibinfo {author} {\bibfnamefont {G.}~\bibnamefont {Montefalcone}}, \bibinfo {author} {\bibfnamefont {S.}~\bibnamefont {Ghosh}}, \bibinfo {author} {\bibfnamefont {K.~K.}\ \bibnamefont {Boddy}}, \bibinfo {author} {\bibfnamefont {D.}~\bibnamefont {Wei Ren~Ho}}, \ and\ \bibinfo {author} {\bibfnamefont {Y.}~\bibnamefont {Tsai}},\ }\href@noop {} {\enquote {\bibinfo {title} {{\it in preparation}},}\ }\BibitemShut {NoStop}%
\bibitem [{\citenamefont {Choi}\ \emph {et~al.}(2018)\citenamefont {Choi}, \citenamefont {Chiang},\ and\ \citenamefont {LoVerde}}]{Choi:2018gho}%
  \BibitemOpen
  \bibfield  {author} {\bibinfo {author} {\bibfnamefont {G.}~\bibnamefont {Choi}}, \bibinfo {author} {\bibfnamefont {C.-T.}\ \bibnamefont {Chiang}}, \ and\ \bibinfo {author} {\bibfnamefont {M.}~\bibnamefont {LoVerde}},\ }\href {\doibase 10.1088/1475-7516/2018/06/044} {\bibfield  {journal} {\bibinfo  {journal} {JCAP}\ }\textbf {\bibinfo {volume} {06}},\ \bibinfo {pages} {044} (\bibinfo {year} {2018})},\ \Eprint {http://arxiv.org/abs/1804.10180} {arXiv:1804.10180 [astro-ph.CO]} \BibitemShut {NoStop}%
\bibitem [{\citenamefont {Baumann}\ \emph {et~al.}(2018)\citenamefont {Baumann}, \citenamefont {Green},\ and\ \citenamefont {Wallisch}}]{Baumann:2017gkg}%
  \BibitemOpen
  \bibfield  {author} {\bibinfo {author} {\bibfnamefont {D.}~\bibnamefont {Baumann}}, \bibinfo {author} {\bibfnamefont {D.}~\bibnamefont {Green}}, \ and\ \bibinfo {author} {\bibfnamefont {B.}~\bibnamefont {Wallisch}},\ }\href {\doibase 10.1088/1475-7516/2018/08/029} {\bibfield  {journal} {\bibinfo  {journal} {JCAP}\ }\textbf {\bibinfo {volume} {08}},\ \bibinfo {pages} {029} (\bibinfo {year} {2018})},\ \Eprint {http://arxiv.org/abs/1712.08067} {arXiv:1712.08067 [astro-ph.CO]} \BibitemShut {NoStop}%
\bibitem [{\citenamefont {Baumann}\ \emph {et~al.}(2019)\citenamefont {Baumann}, \citenamefont {Beutler}, \citenamefont {Flauger}, \citenamefont {Green}, \citenamefont {Slosar}, \citenamefont {Vargas-Maga\~na}, \citenamefont {Wallisch},\ and\ \citenamefont {Y\`{e}che}}]{Baumann:2019keh}%
  \BibitemOpen
  \bibfield  {author} {\bibinfo {author} {\bibfnamefont {D.}~\bibnamefont {Baumann}}, \bibinfo {author} {\bibfnamefont {F.}~\bibnamefont {Beutler}}, \bibinfo {author} {\bibfnamefont {R.}~\bibnamefont {Flauger}}, \bibinfo {author} {\bibfnamefont {D.}~\bibnamefont {Green}}, \bibinfo {author} {\bibfnamefont {A.}~\bibnamefont {Slosar}}, \bibinfo {author} {\bibfnamefont {M.}~\bibnamefont {Vargas-Maga\~na}}, \bibinfo {author} {\bibfnamefont {B.}~\bibnamefont {Wallisch}}, \ and\ \bibinfo {author} {\bibfnamefont {C.}~\bibnamefont {Y\`{e}che}},\ }\href {\doibase 10.1038/s41567-019-0435-6} {\bibfield  {journal} {\bibinfo  {journal} {Nat. Phys.}\ }\textbf {\bibinfo {volume} {15}},\ \bibinfo {pages} {465} (\bibinfo {year} {2019})},\ \Eprint {http://arxiv.org/abs/1803.10741} {arXiv:1803.10741 [astro-ph.CO]} \BibitemShut {NoStop}%
\bibitem [{\citenamefont {Baumann}\ \emph {et~al.}(2017)\citenamefont {Baumann}, \citenamefont {Green},\ and\ \citenamefont {Zaldarriaga}}]{Baumann:2017lmt}%
  \BibitemOpen
  \bibfield  {author} {\bibinfo {author} {\bibfnamefont {D.}~\bibnamefont {Baumann}}, \bibinfo {author} {\bibfnamefont {D.}~\bibnamefont {Green}}, \ and\ \bibinfo {author} {\bibfnamefont {M.}~\bibnamefont {Zaldarriaga}},\ }\href {\doibase 10.1088/1475-7516/2017/11/007} {\bibfield  {journal} {\bibinfo  {journal} {JCAP}\ }\textbf {\bibinfo {volume} {11}},\ \bibinfo {pages} {007} (\bibinfo {year} {2017})},\ \Eprint {http://arxiv.org/abs/1703.00894} {arXiv:1703.00894 [astro-ph.CO]} \BibitemShut {NoStop}%
\bibitem [{\citenamefont {Lee}\ and\ \citenamefont {Hotinli}(2024)}]{Lee:2023uxu}%
  \BibitemOpen
  \bibfield  {author} {\bibinfo {author} {\bibfnamefont {N.}~\bibnamefont {Lee}}\ and\ \bibinfo {author} {\bibfnamefont {S.}~\bibnamefont {Hotinli}},\ }\href {\doibase 10.1103/PhysRevD.109.043502} {\bibfield  {journal} {\bibinfo  {journal} {Phys. Rev. D}\ }\textbf {\bibinfo {volume} {109}},\ \bibinfo {pages} {043502} (\bibinfo {year} {2024})},\ \Eprint {http://arxiv.org/abs/2309.15119} {arXiv:2309.15119 [astro-ph.CO]} \BibitemShut {NoStop}%
\bibitem [{\citenamefont {Dey}\ \emph {et~al.}(2023)\citenamefont {Dey}, \citenamefont {Paul},\ and\ \citenamefont {Pal}}]{Dey:2022ini}%
  \BibitemOpen
  \bibfield  {author} {\bibinfo {author} {\bibfnamefont {A.}~\bibnamefont {Dey}}, \bibinfo {author} {\bibfnamefont {A.}~\bibnamefont {Paul}}, \ and\ \bibinfo {author} {\bibfnamefont {S.}~\bibnamefont {Pal}},\ }\href {\doibase 10.1093/mnras/stad1838} {\bibfield  {journal} {\bibinfo  {journal} {Mon. Not. Roy. Astron. Soc.}\ }\textbf {\bibinfo {volume} {524}},\ \bibinfo {pages} {100} (\bibinfo {year} {2023})},\ \Eprint {http://arxiv.org/abs/2207.02451} {arXiv:2207.02451 [astro-ph.CO]} \BibitemShut {NoStop}%
\bibitem [{\citenamefont {Plombat}\ \emph {et~al.}(2025)\citenamefont {Plombat}, \citenamefont {Simon}, \citenamefont {Flitter},\ and\ \citenamefont {Poulin}}]{Plombat:2024kla}%
  \BibitemOpen
  \bibfield  {author} {\bibinfo {author} {\bibfnamefont {H.}~\bibnamefont {Plombat}}, \bibinfo {author} {\bibfnamefont {T.}~\bibnamefont {Simon}}, \bibinfo {author} {\bibfnamefont {J.}~\bibnamefont {Flitter}}, \ and\ \bibinfo {author} {\bibfnamefont {V.}~\bibnamefont {Poulin}},\ }\href {\doibase 10.1088/1475-7516/2025/01/071} {\bibfield  {journal} {\bibinfo  {journal} {JCAP}\ }\textbf {\bibinfo {volume} {01}},\ \bibinfo {pages} {071} (\bibinfo {year} {2025})},\ \Eprint {http://arxiv.org/abs/2410.01486} {arXiv:2410.01486 [astro-ph.CO]} \BibitemShut {NoStop}%
\bibitem [{\citenamefont {Dhuria}\ and\ \citenamefont {Teli}(2024)}]{Dhuria:2024zwh}%
  \BibitemOpen
  \bibfield  {author} {\bibinfo {author} {\bibfnamefont {M.}~\bibnamefont {Dhuria}}\ and\ \bibinfo {author} {\bibfnamefont {B.~G.}\ \bibnamefont {Teli}},\ }\href {\doibase 10.1103/PhysRevD.110.123033} {\bibfield  {journal} {\bibinfo  {journal} {Phys. Rev. D}\ }\textbf {\bibinfo {volume} {110}},\ \bibinfo {pages} {123033} (\bibinfo {year} {2024})},\ \Eprint {http://arxiv.org/abs/2406.19279} {arXiv:2406.19279 [hep-ph]} \BibitemShut {NoStop}%
\bibitem [{\citenamefont {Libanore}\ \emph {et~al.}(2025)\citenamefont {Libanore}, \citenamefont {Ghosh}, \citenamefont {Kovetz}, \citenamefont {Boddy},\ and\ \citenamefont {Raccanelli}}]{Libanore:2025ack}%
  \BibitemOpen
  \bibfield  {author} {\bibinfo {author} {\bibfnamefont {S.}~\bibnamefont {Libanore}}, \bibinfo {author} {\bibfnamefont {S.}~\bibnamefont {Ghosh}}, \bibinfo {author} {\bibfnamefont {E.~D.}\ \bibnamefont {Kovetz}}, \bibinfo {author} {\bibfnamefont {K.~K.}\ \bibnamefont {Boddy}}, \ and\ \bibinfo {author} {\bibfnamefont {A.}~\bibnamefont {Raccanelli}},\ }\href@noop {} {\enquote {\bibinfo {title} {{Joint 21-cm and CMB Forecasts for Constraining Self-Interacting Massive Neutrinos}},}\ } (\bibinfo {year} {2025}),\ \Eprint {http://arxiv.org/abs/2504.15348} {arXiv:2504.15348 [astro-ph.CO]} \BibitemShut {NoStop}%
\bibitem [{\citenamefont {{Mu{\~n}oz}}(2023)}]{munoz23}%
  \BibitemOpen
  \bibfield  {author} {\bibinfo {author} {\bibfnamefont {J.~B.}\ \bibnamefont {{Mu{\~n}oz}}},\ }\href {\doibase 10.1093/mnras/stad1512} {\bibfield  {journal} {\bibinfo  {journal} {\mnras}\ }\textbf {\bibinfo {volume} {523}},\ \bibinfo {pages} {2587} (\bibinfo {year} {2023})},\ \Eprint {http://arxiv.org/abs/2302.08506} {arXiv:2302.08506 [astro-ph.CO]} \BibitemShut {NoStop}%
\bibitem [{\citenamefont {{Mu{\~n}oz}}\ \emph {et~al.}(2023)\citenamefont {{Mu{\~n}oz}}, \citenamefont {{Mirocha}}, \citenamefont {{Furlanetto}},\ and\ \citenamefont {{Sabti}}}]{munoz23b}%
  \BibitemOpen
  \bibfield  {author} {\bibinfo {author} {\bibfnamefont {J.~B.}\ \bibnamefont {{Mu{\~n}oz}}}, \bibinfo {author} {\bibfnamefont {J.}~\bibnamefont {{Mirocha}}}, \bibinfo {author} {\bibfnamefont {S.}~\bibnamefont {{Furlanetto}}}, \ and\ \bibinfo {author} {\bibfnamefont {N.}~\bibnamefont {{Sabti}}},\ }\href {\doibase 10.1093/mnrasl/slad115} {\bibfield  {journal} {\bibinfo  {journal} {\mnras}\ }\textbf {\bibinfo {volume} {526}},\ \bibinfo {pages} {L47} (\bibinfo {year} {2023})},\ \Eprint {http://arxiv.org/abs/2306.09403} {arXiv:2306.09403 [astro-ph.CO]} \BibitemShut {NoStop}%
\bibitem [{\citenamefont {{Furlanetto}}\ \emph {et~al.}(2006)\citenamefont {{Furlanetto}}, \citenamefont {{Oh}},\ and\ \citenamefont {{Briggs}}}]{Furlanetto:2006jb}%
  \BibitemOpen
  \bibfield  {author} {\bibinfo {author} {\bibfnamefont {S.~R.}\ \bibnamefont {{Furlanetto}}}, \bibinfo {author} {\bibfnamefont {S.~P.}\ \bibnamefont {{Oh}}}, \ and\ \bibinfo {author} {\bibfnamefont {F.~H.}\ \bibnamefont {{Briggs}}},\ }\href {\doibase 10.1016/j.physrep.2006.08.002} {\bibfield  {journal} {\bibinfo  {journal} {\physrep}\ }\textbf {\bibinfo {volume} {433}},\ \bibinfo {pages} {181} (\bibinfo {year} {2006})},\ \Eprint {http://arxiv.org/abs/astro-ph/0608032} {arXiv:astro-ph/0608032 [astro-ph]} \BibitemShut {NoStop}%
\bibitem [{\citenamefont {Pritchard}\ and\ \citenamefont {Loeb}(2008)}]{Pritchard:2008da}%
  \BibitemOpen
  \bibfield  {author} {\bibinfo {author} {\bibfnamefont {J.~R.}\ \bibnamefont {Pritchard}}\ and\ \bibinfo {author} {\bibfnamefont {A.}~\bibnamefont {Loeb}},\ }\href {\doibase 10.1103/PhysRevD.78.103511} {\bibfield  {journal} {\bibinfo  {journal} {Phys. Rev. D}\ }\textbf {\bibinfo {volume} {78}},\ \bibinfo {pages} {103511} (\bibinfo {year} {2008})},\ \Eprint {http://arxiv.org/abs/0802.2102} {arXiv:0802.2102 [astro-ph]} \BibitemShut {NoStop}%
\bibitem [{\citenamefont {Pritchard}\ and\ \citenamefont {Loeb}(2010)}]{Pritchard:2010pa}%
  \BibitemOpen
  \bibfield  {author} {\bibinfo {author} {\bibfnamefont {J.~R.}\ \bibnamefont {Pritchard}}\ and\ \bibinfo {author} {\bibfnamefont {A.}~\bibnamefont {Loeb}},\ }\href {\doibase 10.1103/PhysRevD.82.023006} {\bibfield  {journal} {\bibinfo  {journal} {Phys. Rev. D}\ }\textbf {\bibinfo {volume} {82}},\ \bibinfo {pages} {023006} (\bibinfo {year} {2010})},\ \Eprint {http://arxiv.org/abs/1005.4057} {arXiv:1005.4057 [astro-ph.CO]} \BibitemShut {NoStop}%
\bibitem [{\citenamefont {{Pritchard}}\ and\ \citenamefont {{Loeb}}(2012)}]{pritchard12}%
  \BibitemOpen
  \bibfield  {author} {\bibinfo {author} {\bibfnamefont {J.~R.}\ \bibnamefont {{Pritchard}}}\ and\ \bibinfo {author} {\bibfnamefont {A.}~\bibnamefont {{Loeb}}},\ }\href {\doibase 10.1088/0034-4885/75/8/086901} {\bibfield  {journal} {\bibinfo  {journal} {Reports on Progress in Physics}\ }\textbf {\bibinfo {volume} {75}},\ \bibinfo {eid} {086901} (\bibinfo {year} {2012})},\ \Eprint {http://arxiv.org/abs/1109.6012} {arXiv:1109.6012 [astro-ph.CO]} \BibitemShut {NoStop}%
\bibitem [{\citenamefont {Flitter}\ \emph {et~al.}(2025)\citenamefont {Flitter}, \citenamefont {Libanore},\ and\ \citenamefont {Kovetz}}]{Flitter:2024eay}%
  \BibitemOpen
  \bibfield  {author} {\bibinfo {author} {\bibfnamefont {J.}~\bibnamefont {Flitter}}, \bibinfo {author} {\bibfnamefont {S.}~\bibnamefont {Libanore}}, \ and\ \bibinfo {author} {\bibfnamefont {E.~D.}\ \bibnamefont {Kovetz}},\ }\href {\doibase 10.1103/zc4v-2yx4} {\bibfield  {journal} {\bibinfo  {journal} {Phys. Rev. D}\ }\textbf {\bibinfo {volume} {112}},\ \bibinfo {pages} {023537} (\bibinfo {year} {2025})},\ \Eprint {http://arxiv.org/abs/2411.00089} {arXiv:2411.00089 [astro-ph.CO]} \BibitemShut {NoStop}%
\bibitem [{\citenamefont {{N. Aghanim \textit{et al.} (Planck Collaboration)}}(2020)}]{Planck:2018vyg}%
  \BibitemOpen
  \bibfield  {author} {\bibinfo {author} {\bibnamefont {{N. Aghanim \textit{et al.} (Planck Collaboration)}}},\ }\href {\doibase 10.1051/0004-6361/201833910} {\bibfield  {journal} {\bibinfo  {journal} {Astron. Astrophys.}\ }\textbf {\bibinfo {volume} {641}},\ \bibinfo {pages} {A6} (\bibinfo {year} {2020})},\ \Eprint {http://arxiv.org/abs/1807.06209} {arXiv:1807.06209 [astro-ph.CO]} \BibitemShut {NoStop}%
\bibitem [{\citenamefont {Whitford}\ \emph {et~al.}(2024)\citenamefont {Whitford} \emph {et~al.}}]{Whitford:2024ecj}%
  \BibitemOpen
  \bibfield  {author} {\bibinfo {author} {\bibfnamefont {A.~M.}\ \bibnamefont {Whitford}} \emph {et~al.},\ }\href@noop {} {\  (\bibinfo {year} {2024})},\ \Eprint {http://arxiv.org/abs/2412.05990} {arXiv:2412.05990 [astro-ph.CO]} \BibitemShut {NoStop}%
\bibitem [{\citenamefont {Saravanan}\ \emph {et~al.}(2025)\citenamefont {Saravanan}, \citenamefont {Brinckmann}, \citenamefont {Loverde},\ and\ \citenamefont {Weiner}}]{Saravanan:2025cyi}%
  \BibitemOpen
  \bibfield  {author} {\bibinfo {author} {\bibfnamefont {M.~M.}\ \bibnamefont {Saravanan}}, \bibinfo {author} {\bibfnamefont {T.}~\bibnamefont {Brinckmann}}, \bibinfo {author} {\bibfnamefont {M.}~\bibnamefont {Loverde}}, \ and\ \bibinfo {author} {\bibfnamefont {Z.~J.}\ \bibnamefont {Weiner}},\ }\href@noop {} {\  (\bibinfo {year} {2025})},\ \Eprint {http://arxiv.org/abs/2503.04671} {arXiv:2503.04671 [astro-ph.CO]} \BibitemShut {NoStop}%
\bibitem [{\citenamefont {Cruz}\ \emph {et~al.}(2025)\citenamefont {Cruz}, \citenamefont {Munoz}, \citenamefont {Sabti},\ and\ \citenamefont {Kamionkowski}}]{Cruz:2024fsv}%
  \BibitemOpen
  \bibfield  {author} {\bibinfo {author} {\bibfnamefont {H.~A.~G.}\ \bibnamefont {Cruz}}, \bibinfo {author} {\bibfnamefont {J.~B.}\ \bibnamefont {Munoz}}, \bibinfo {author} {\bibfnamefont {N.}~\bibnamefont {Sabti}}, \ and\ \bibinfo {author} {\bibfnamefont {M.}~\bibnamefont {Kamionkowski}},\ }\href {\doibase 10.1103/PhysRevD.111.083503} {\bibfield  {journal} {\bibinfo  {journal} {Phys. Rev. D}\ }\textbf {\bibinfo {volume} {111}},\ \bibinfo {pages} {083503} (\bibinfo {year} {2025})},\ \Eprint {http://arxiv.org/abs/2407.18294} {arXiv:2407.18294 [astro-ph.CO]} \BibitemShut {NoStop}%
\bibitem [{\citenamefont {{Barkana}}\ and\ \citenamefont {{Loeb}}(2006)}]{barkana06}%
  \BibitemOpen
  \bibfield  {author} {\bibinfo {author} {\bibfnamefont {R.}~\bibnamefont {{Barkana}}}\ and\ \bibinfo {author} {\bibfnamefont {A.}~\bibnamefont {{Loeb}}},\ }\href {\doibase 10.1111/j.1745-3933.2006.00222.x} {\bibfield  {journal} {\bibinfo  {journal} {\mnras}\ }\textbf {\bibinfo {volume} {372}},\ \bibinfo {pages} {L43} (\bibinfo {year} {2006})},\ \Eprint {http://arxiv.org/abs/astro-ph/0512453} {arXiv:astro-ph/0512453 [astro-ph]} \BibitemShut {NoStop}%
\bibitem [{\citenamefont {{Mao}}\ \emph {et~al.}(2012)\citenamefont {{Mao}}, \citenamefont {{Shapiro}}, \citenamefont {{Mellema}}, \citenamefont {{Iliev}}, \citenamefont {{Koda}},\ and\ \citenamefont {{Ahn}}}]{mao12}%
  \BibitemOpen
  \bibfield  {author} {\bibinfo {author} {\bibfnamefont {Y.}~\bibnamefont {{Mao}}}, \bibinfo {author} {\bibfnamefont {P.~R.}\ \bibnamefont {{Shapiro}}}, \bibinfo {author} {\bibfnamefont {G.}~\bibnamefont {{Mellema}}}, \bibinfo {author} {\bibfnamefont {I.~T.}\ \bibnamefont {{Iliev}}}, \bibinfo {author} {\bibfnamefont {J.}~\bibnamefont {{Koda}}}, \ and\ \bibinfo {author} {\bibfnamefont {K.}~\bibnamefont {{Ahn}}},\ }\href {\doibase 10.1111/j.1365-2966.2012.20471.x} {\bibfield  {journal} {\bibinfo  {journal} {\mnras}\ }\textbf {\bibinfo {volume} {422}},\ \bibinfo {pages} {926} (\bibinfo {year} {2012})},\ \Eprint {http://arxiv.org/abs/1104.2094} {arXiv:1104.2094 [astro-ph.CO]} \BibitemShut {NoStop}%
\bibitem [{\citenamefont {{Wouthuysen}}(1952)}]{wouthuysen52}%
  \BibitemOpen
  \bibfield  {author} {\bibinfo {author} {\bibfnamefont {S.~A.}\ \bibnamefont {{Wouthuysen}}},\ }\href {\doibase 10.1086/106661} {\bibfield  {journal} {\bibinfo  {journal} {\aj}\ }\textbf {\bibinfo {volume} {57}},\ \bibinfo {pages} {31} (\bibinfo {year} {1952})}\BibitemShut {NoStop}%
\bibitem [{\citenamefont {{Field}}(1959)}]{field59}%
  \BibitemOpen
  \bibfield  {author} {\bibinfo {author} {\bibfnamefont {G.~B.}\ \bibnamefont {{Field}}},\ }\href {\doibase 10.1086/146653} {\bibfield  {journal} {\bibinfo  {journal} {\apj}\ }\textbf {\bibinfo {volume} {129}},\ \bibinfo {pages} {536} (\bibinfo {year} {1959})}\BibitemShut {NoStop}%
\bibitem [{\citenamefont {{Hirata}}(2006)}]{hirata06}%
  \BibitemOpen
  \bibfield  {author} {\bibinfo {author} {\bibfnamefont {C.~M.}\ \bibnamefont {{Hirata}}},\ }\href {\doibase 10.1111/j.1365-2966.2005.09949.x} {\bibfield  {journal} {\bibinfo  {journal} {\mnras}\ }\textbf {\bibinfo {volume} {367}},\ \bibinfo {pages} {259} (\bibinfo {year} {2006})},\ \Eprint {http://arxiv.org/abs/astro-ph/0507102} {arXiv:astro-ph/0507102 [astro-ph]} \BibitemShut {NoStop}%
\bibitem [{\citenamefont {{Tseliakhovich}}\ and\ \citenamefont {{Hirata}}(2010)}]{tseliakhovich10}%
  \BibitemOpen
  \bibfield  {author} {\bibinfo {author} {\bibfnamefont {D.}~\bibnamefont {{Tseliakhovich}}}\ and\ \bibinfo {author} {\bibfnamefont {C.}~\bibnamefont {{Hirata}}},\ }\href {\doibase 10.1103/PhysRevD.82.083520} {\bibfield  {journal} {\bibinfo  {journal} {\prd}\ }\textbf {\bibinfo {volume} {82}},\ \bibinfo {eid} {083520} (\bibinfo {year} {2010})},\ \Eprint {http://arxiv.org/abs/1005.2416} {arXiv:1005.2416 [astro-ph.CO]} \BibitemShut {NoStop}%
\bibitem [{\citenamefont {{Dalal}}\ \emph {et~al.}(2010)\citenamefont {{Dalal}}, \citenamefont {{Pen}},\ and\ \citenamefont {{Seljak}}}]{dalal10}%
  \BibitemOpen
  \bibfield  {author} {\bibinfo {author} {\bibfnamefont {N.}~\bibnamefont {{Dalal}}}, \bibinfo {author} {\bibfnamefont {U.-L.}\ \bibnamefont {{Pen}}}, \ and\ \bibinfo {author} {\bibfnamefont {U.}~\bibnamefont {{Seljak}}},\ }\href {\doibase 10.1088/1475-7516/2010/11/007} {\bibfield  {journal} {\bibinfo  {journal} {\jcap}\ }\textbf {\bibinfo {volume} {2010}},\ \bibinfo {eid} {007} (\bibinfo {year} {2010})},\ \Eprint {http://arxiv.org/abs/1009.4704} {arXiv:1009.4704 [astro-ph.CO]} \BibitemShut {NoStop}%
\bibitem [{\citenamefont {{Hamann}}\ \emph {et~al.}(2010)\citenamefont {{Hamann}}, \citenamefont {{Hannestad}}, \citenamefont {{Lesgourgues}}, \citenamefont {{Rampf}},\ and\ \citenamefont {{Wong}}}]{Hamann:2010pw}%
  \BibitemOpen
  \bibfield  {author} {\bibinfo {author} {\bibfnamefont {J.}~\bibnamefont {{Hamann}}}, \bibinfo {author} {\bibfnamefont {S.}~\bibnamefont {{Hannestad}}}, \bibinfo {author} {\bibfnamefont {J.}~\bibnamefont {{Lesgourgues}}}, \bibinfo {author} {\bibfnamefont {C.}~\bibnamefont {{Rampf}}}, \ and\ \bibinfo {author} {\bibfnamefont {Y.~Y.~Y.}\ \bibnamefont {{Wong}}},\ }\href {\doibase 10.1088/1475-7516/2010/07/022} {\bibfield  {journal} {\bibinfo  {journal} {\jcap}\ }\textbf {\bibinfo {volume} {2010}},\ \bibinfo {eid} {022} (\bibinfo {year} {2010})},\ \Eprint {http://arxiv.org/abs/1003.3999} {arXiv:1003.3999 [astro-ph.CO]} \BibitemShut {NoStop}%
\bibitem [{\citenamefont {{Naoz}}\ \emph {et~al.}(2012)\citenamefont {{Naoz}}, \citenamefont {{Yoshida}},\ and\ \citenamefont {{Gnedin}}}]{naoz12}%
  \BibitemOpen
  \bibfield  {author} {\bibinfo {author} {\bibfnamefont {S.}~\bibnamefont {{Naoz}}}, \bibinfo {author} {\bibfnamefont {N.}~\bibnamefont {{Yoshida}}}, \ and\ \bibinfo {author} {\bibfnamefont {N.~Y.}\ \bibnamefont {{Gnedin}}},\ }\href {\doibase 10.1088/0004-637X/747/2/128} {\bibfield  {journal} {\bibinfo  {journal} {\apj}\ }\textbf {\bibinfo {volume} {747}},\ \bibinfo {eid} {128} (\bibinfo {year} {2012})},\ \Eprint {http://arxiv.org/abs/1108.5176} {arXiv:1108.5176 [astro-ph.CO]} \BibitemShut {NoStop}%
\bibitem [{\citenamefont {{Bovy}}\ and\ \citenamefont {{Dvorkin}}(2013)}]{bovy13}%
  \BibitemOpen
  \bibfield  {author} {\bibinfo {author} {\bibfnamefont {J.}~\bibnamefont {{Bovy}}}\ and\ \bibinfo {author} {\bibfnamefont {C.}~\bibnamefont {{Dvorkin}}},\ }\href {\doibase 10.1088/0004-637X/768/1/70} {\bibfield  {journal} {\bibinfo  {journal} {\apj}\ }\textbf {\bibinfo {volume} {768}},\ \bibinfo {eid} {70} (\bibinfo {year} {2013})},\ \Eprint {http://arxiv.org/abs/1205.2083} {arXiv:1205.2083 [astro-ph.CO]} \BibitemShut {NoStop}%
\bibitem [{\citenamefont {{Greif}}\ \emph {et~al.}(2011)\citenamefont {{Greif}}, \citenamefont {{White}}, \citenamefont {{Klessen}},\ and\ \citenamefont {{Springel}}}]{greif11}%
  \BibitemOpen
  \bibfield  {author} {\bibinfo {author} {\bibfnamefont {T.~H.}\ \bibnamefont {{Greif}}}, \bibinfo {author} {\bibfnamefont {S.~D.~M.}\ \bibnamefont {{White}}}, \bibinfo {author} {\bibfnamefont {R.~S.}\ \bibnamefont {{Klessen}}}, \ and\ \bibinfo {author} {\bibfnamefont {V.}~\bibnamefont {{Springel}}},\ }\href {\doibase 10.1088/0004-637X/736/2/147} {\bibfield  {journal} {\bibinfo  {journal} {\apj}\ }\textbf {\bibinfo {volume} {736}},\ \bibinfo {eid} {147} (\bibinfo {year} {2011})},\ \Eprint {http://arxiv.org/abs/1101.5493} {arXiv:1101.5493 [astro-ph.CO]} \BibitemShut {NoStop}%
\bibitem [{\citenamefont {{Hirano}}\ \emph {et~al.}(2018)\citenamefont {{Hirano}}, \citenamefont {{Yoshida}}, \citenamefont {{Sakurai}},\ and\ \citenamefont {{Fujii}}}]{hirano18}%
  \BibitemOpen
  \bibfield  {author} {\bibinfo {author} {\bibfnamefont {S.}~\bibnamefont {{Hirano}}}, \bibinfo {author} {\bibfnamefont {N.}~\bibnamefont {{Yoshida}}}, \bibinfo {author} {\bibfnamefont {Y.}~\bibnamefont {{Sakurai}}}, \ and\ \bibinfo {author} {\bibfnamefont {M.~S.}\ \bibnamefont {{Fujii}}},\ }\href {\doibase 10.3847/1538-4357/aaaaba} {\bibfield  {journal} {\bibinfo  {journal} {\apj}\ }\textbf {\bibinfo {volume} {855}},\ \bibinfo {eid} {17} (\bibinfo {year} {2018})},\ \Eprint {http://arxiv.org/abs/1711.07315} {arXiv:1711.07315 [astro-ph.GA]} \BibitemShut {NoStop}%
\bibitem [{\citenamefont {{Tseliakhovich}}\ \emph {et~al.}(2011)\citenamefont {{Tseliakhovich}}, \citenamefont {{Barkana}},\ and\ \citenamefont {{Hirata}}}]{tseliakhovich11}%
  \BibitemOpen
  \bibfield  {author} {\bibinfo {author} {\bibfnamefont {D.}~\bibnamefont {{Tseliakhovich}}}, \bibinfo {author} {\bibfnamefont {R.}~\bibnamefont {{Barkana}}}, \ and\ \bibinfo {author} {\bibfnamefont {C.~M.}\ \bibnamefont {{Hirata}}},\ }\href {\doibase 10.1111/j.1365-2966.2011.19541.x} {\bibfield  {journal} {\bibinfo  {journal} {\mnras}\ }\textbf {\bibinfo {volume} {418}},\ \bibinfo {pages} {906} (\bibinfo {year} {2011})},\ \Eprint {http://arxiv.org/abs/1012.2574} {arXiv:1012.2574 [astro-ph.CO]} \BibitemShut {NoStop}%
\bibitem [{\citenamefont {{Stacy}}\ \emph {et~al.}(2011)\citenamefont {{Stacy}}, \citenamefont {{Bromm}},\ and\ \citenamefont {{Loeb}}}]{stacy11}%
  \BibitemOpen
  \bibfield  {author} {\bibinfo {author} {\bibfnamefont {A.}~\bibnamefont {{Stacy}}}, \bibinfo {author} {\bibfnamefont {V.}~\bibnamefont {{Bromm}}}, \ and\ \bibinfo {author} {\bibfnamefont {A.}~\bibnamefont {{Loeb}}},\ }\href {\doibase 10.1088/2041-8205/730/1/L1} {\bibfield  {journal} {\bibinfo  {journal} {\apjl}\ }\textbf {\bibinfo {volume} {730}},\ \bibinfo {eid} {L1} (\bibinfo {year} {2011})},\ \Eprint {http://arxiv.org/abs/1011.4512} {arXiv:1011.4512 [astro-ph.CO]} \BibitemShut {NoStop}%
\bibitem [{\citenamefont {{O'Leary}}\ and\ \citenamefont {{McQuinn}}(2012)}]{oleary12}%
  \BibitemOpen
  \bibfield  {author} {\bibinfo {author} {\bibfnamefont {R.~M.}\ \bibnamefont {{O'Leary}}}\ and\ \bibinfo {author} {\bibfnamefont {M.}~\bibnamefont {{McQuinn}}},\ }\href {\doibase 10.1088/0004-637X/760/1/4} {\bibfield  {journal} {\bibinfo  {journal} {\apj}\ }\textbf {\bibinfo {volume} {760}},\ \bibinfo {eid} {4} (\bibinfo {year} {2012})},\ \Eprint {http://arxiv.org/abs/1204.1344} {arXiv:1204.1344 [astro-ph.CO]} \BibitemShut {NoStop}%
\bibitem [{\citenamefont {{Naoz}}\ \emph {et~al.}(2013)\citenamefont {{Naoz}}, \citenamefont {{Yoshida}},\ and\ \citenamefont {{Gnedin}}}]{naoz13}%
  \BibitemOpen
  \bibfield  {author} {\bibinfo {author} {\bibfnamefont {S.}~\bibnamefont {{Naoz}}}, \bibinfo {author} {\bibfnamefont {N.}~\bibnamefont {{Yoshida}}}, \ and\ \bibinfo {author} {\bibfnamefont {N.~Y.}\ \bibnamefont {{Gnedin}}},\ }\href {\doibase 10.1088/0004-637X/763/1/27} {\bibfield  {journal} {\bibinfo  {journal} {\apj}\ }\textbf {\bibinfo {volume} {763}},\ \bibinfo {eid} {27} (\bibinfo {year} {2013})},\ \Eprint {http://arxiv.org/abs/1207.5515} {arXiv:1207.5515 [astro-ph.CO]} \BibitemShut {NoStop}%
\bibitem [{\citenamefont {{Ferraro}}\ \emph {et~al.}(2012)\citenamefont {{Ferraro}}, \citenamefont {{Smith}},\ and\ \citenamefont {{Dvorkin}}}]{ferraro12}%
  \BibitemOpen
  \bibfield  {author} {\bibinfo {author} {\bibfnamefont {S.}~\bibnamefont {{Ferraro}}}, \bibinfo {author} {\bibfnamefont {K.~M.}\ \bibnamefont {{Smith}}}, \ and\ \bibinfo {author} {\bibfnamefont {C.}~\bibnamefont {{Dvorkin}}},\ }\href {\doibase 10.1103/PhysRevD.85.043523} {\bibfield  {journal} {\bibinfo  {journal} {\prd}\ }\textbf {\bibinfo {volume} {85}},\ \bibinfo {eid} {043523} (\bibinfo {year} {2012})},\ \Eprint {http://arxiv.org/abs/1110.2182} {arXiv:1110.2182 [astro-ph.CO]} \BibitemShut {NoStop}%
\bibitem [{\citenamefont {{Yoo}}\ \emph {et~al.}(2011)\citenamefont {{Yoo}}, \citenamefont {{Dalal}},\ and\ \citenamefont {{Seljak}}}]{Yoo:2011tq}%
  \BibitemOpen
  \bibfield  {author} {\bibinfo {author} {\bibfnamefont {J.}~\bibnamefont {{Yoo}}}, \bibinfo {author} {\bibfnamefont {N.}~\bibnamefont {{Dalal}}}, \ and\ \bibinfo {author} {\bibfnamefont {U.}~\bibnamefont {{Seljak}}},\ }\href {\doibase 10.1088/1475-7516/2011/07/018} {\bibfield  {journal} {\bibinfo  {journal} {\jcap}\ }\textbf {\bibinfo {volume} {2011}},\ \bibinfo {eid} {018} (\bibinfo {year} {2011})},\ \Eprint {http://arxiv.org/abs/1105.3732} {arXiv:1105.3732 [astro-ph.CO]} \BibitemShut {NoStop}%
\bibitem [{\citenamefont {{Mu{\~n}oz}}\ \emph {et~al.}(2022)\citenamefont {{Mu{\~n}oz}}, \citenamefont {{Qin}}, \citenamefont {{Mesinger}}, \citenamefont {{Murray}}, \citenamefont {{Greig}},\ and\ \citenamefont {{Mason}}}]{munoz22}%
  \BibitemOpen
  \bibfield  {author} {\bibinfo {author} {\bibfnamefont {J.~B.}\ \bibnamefont {{Mu{\~n}oz}}}, \bibinfo {author} {\bibfnamefont {Y.}~\bibnamefont {{Qin}}}, \bibinfo {author} {\bibfnamefont {A.}~\bibnamefont {{Mesinger}}}, \bibinfo {author} {\bibfnamefont {S.~G.}\ \bibnamefont {{Murray}}}, \bibinfo {author} {\bibfnamefont {B.}~\bibnamefont {{Greig}}}, \ and\ \bibinfo {author} {\bibfnamefont {C.}~\bibnamefont {{Mason}}},\ }\href {\doibase 10.1093/mnras/stac185} {\bibfield  {journal} {\bibinfo  {journal} {\mnras}\ }\textbf {\bibinfo {volume} {511}},\ \bibinfo {pages} {3657} (\bibinfo {year} {2022})},\ \Eprint {http://arxiv.org/abs/2110.13919} {arXiv:2110.13919 [astro-ph.CO]} \BibitemShut {NoStop}%
\bibitem [{\citenamefont {{Ali-Ha{\"\i}moud}}\ \emph {et~al.}(2014)\citenamefont {{Ali-Ha{\"\i}moud}}, \citenamefont {{Meerburg}},\ and\ \citenamefont {{Yuan}}}]{alihaimoud14}%
  \BibitemOpen
  \bibfield  {author} {\bibinfo {author} {\bibfnamefont {Y.}~\bibnamefont {{Ali-Ha{\"\i}moud}}}, \bibinfo {author} {\bibfnamefont {P.~D.}\ \bibnamefont {{Meerburg}}}, \ and\ \bibinfo {author} {\bibfnamefont {S.}~\bibnamefont {{Yuan}}},\ }\href {\doibase 10.1103/PhysRevD.89.083506} {\bibfield  {journal} {\bibinfo  {journal} {\prd}\ }\textbf {\bibinfo {volume} {89}},\ \bibinfo {eid} {083506} (\bibinfo {year} {2014})},\ \Eprint {http://arxiv.org/abs/1312.4948} {arXiv:1312.4948 [astro-ph.CO]} \BibitemShut {NoStop}%
\bibitem [{\citenamefont {Mu{\~n}oz}\ \emph {et~al.}(2018)\citenamefont {Mu{\~n}oz}, \citenamefont {Dvorkin},\ and\ \citenamefont {Loeb}}]{Munoz:2018jwq}%
  \BibitemOpen
  \bibfield  {author} {\bibinfo {author} {\bibfnamefont {J.~B.}\ \bibnamefont {Mu{\~n}oz}}, \bibinfo {author} {\bibfnamefont {C.}~\bibnamefont {Dvorkin}}, \ and\ \bibinfo {author} {\bibfnamefont {A.}~\bibnamefont {Loeb}},\ }\href {\doibase 10.1103/PhysRevLett.121.121301} {\bibfield  {journal} {\bibinfo  {journal} {Phys. Rev. Lett.}\ }\textbf {\bibinfo {volume} {121}},\ \bibinfo {pages} {121301} (\bibinfo {year} {2018})},\ \Eprint {http://arxiv.org/abs/1804.01092} {arXiv:1804.01092 [astro-ph.CO]} \BibitemShut {NoStop}%
\bibitem [{\citenamefont {{Wyithe}}\ \emph {et~al.}(2008)\citenamefont {{Wyithe}}, \citenamefont {{Loeb}},\ and\ \citenamefont {{Geil}}}]{wyithe08}%
  \BibitemOpen
  \bibfield  {author} {\bibinfo {author} {\bibfnamefont {J.~S.~B.}\ \bibnamefont {{Wyithe}}}, \bibinfo {author} {\bibfnamefont {A.}~\bibnamefont {{Loeb}}}, \ and\ \bibinfo {author} {\bibfnamefont {P.~M.}\ \bibnamefont {{Geil}}},\ }\href {\doibase 10.1111/j.1365-2966.2007.12631.x} {\bibfield  {journal} {\bibinfo  {journal} {\mnras}\ }\textbf {\bibinfo {volume} {383}},\ \bibinfo {pages} {1195} (\bibinfo {year} {2008})},\ \Eprint {http://arxiv.org/abs/0709.2955} {arXiv:0709.2955 [astro-ph]} \BibitemShut {NoStop}%
\bibitem [{\citenamefont {{Wyithe}}\ and\ \citenamefont {{Loeb}}(2008)}]{wyithe08b}%
  \BibitemOpen
  \bibfield  {author} {\bibinfo {author} {\bibfnamefont {J.~S.~B.}\ \bibnamefont {{Wyithe}}}\ and\ \bibinfo {author} {\bibfnamefont {A.}~\bibnamefont {{Loeb}}},\ }\href {\doibase 10.1111/j.1365-2966.2007.12568.x} {\bibfield  {journal} {\bibinfo  {journal} {\mnras}\ }\textbf {\bibinfo {volume} {383}},\ \bibinfo {pages} {606} (\bibinfo {year} {2008})},\ \Eprint {http://arxiv.org/abs/0708.3392} {arXiv:0708.3392 [astro-ph]} \BibitemShut {NoStop}%
\bibitem [{\citenamefont {{Loeb}}\ and\ \citenamefont {{Wyithe}}(2008)}]{loeb08}%
  \BibitemOpen
  \bibfield  {author} {\bibinfo {author} {\bibfnamefont {A.}~\bibnamefont {{Loeb}}}\ and\ \bibinfo {author} {\bibfnamefont {J.~S.~B.}\ \bibnamefont {{Wyithe}}},\ }\href {\doibase 10.1103/PhysRevLett.100.161301} {\bibfield  {journal} {\bibinfo  {journal} {\prl}\ }\textbf {\bibinfo {volume} {100}},\ \bibinfo {eid} {161301} (\bibinfo {year} {2008})},\ \Eprint {http://arxiv.org/abs/0801.1677} {arXiv:0801.1677 [astro-ph]} \BibitemShut {NoStop}%
\bibitem [{\citenamefont {{Chang}}\ \emph {et~al.}(2008)\citenamefont {{Chang}}, \citenamefont {{Pen}}, \citenamefont {{Peterson}},\ and\ \citenamefont {{McDonald}}}]{chang08}%
  \BibitemOpen
  \bibfield  {author} {\bibinfo {author} {\bibfnamefont {T.-C.}\ \bibnamefont {{Chang}}}, \bibinfo {author} {\bibfnamefont {U.-L.}\ \bibnamefont {{Pen}}}, \bibinfo {author} {\bibfnamefont {J.~B.}\ \bibnamefont {{Peterson}}}, \ and\ \bibinfo {author} {\bibfnamefont {P.}~\bibnamefont {{McDonald}}},\ }\href {\doibase 10.1103/PhysRevLett.100.091303} {\bibfield  {journal} {\bibinfo  {journal} {\prl}\ }\textbf {\bibinfo {volume} {100}},\ \bibinfo {eid} {091303} (\bibinfo {year} {2008})},\ \Eprint {http://arxiv.org/abs/0709.3672} {arXiv:0709.3672 [astro-ph]} \BibitemShut {NoStop}%
\bibitem [{\citenamefont {{Seo}}\ \emph {et~al.}(2010)\citenamefont {{Seo}}, \citenamefont {{Dodelson}}, \citenamefont {{Marriner}}, \citenamefont {{Mcginnis}}, \citenamefont {{Stebbins}}, \citenamefont {{Stoughton}},\ and\ \citenamefont {{Vallinotto}}}]{seo10}%
  \BibitemOpen
  \bibfield  {author} {\bibinfo {author} {\bibfnamefont {H.-J.}\ \bibnamefont {{Seo}}}, \bibinfo {author} {\bibfnamefont {S.}~\bibnamefont {{Dodelson}}}, \bibinfo {author} {\bibfnamefont {J.}~\bibnamefont {{Marriner}}}, \bibinfo {author} {\bibfnamefont {D.}~\bibnamefont {{Mcginnis}}}, \bibinfo {author} {\bibfnamefont {A.}~\bibnamefont {{Stebbins}}}, \bibinfo {author} {\bibfnamefont {C.}~\bibnamefont {{Stoughton}}}, \ and\ \bibinfo {author} {\bibfnamefont {A.}~\bibnamefont {{Vallinotto}}},\ }\href {\doibase 10.1088/0004-637X/721/1/164} {\bibfield  {journal} {\bibinfo  {journal} {\apj}\ }\textbf {\bibinfo {volume} {721}},\ \bibinfo {pages} {164} (\bibinfo {year} {2010})},\ \Eprint {http://arxiv.org/abs/0910.5007} {arXiv:0910.5007 [astro-ph.CO]} \BibitemShut {NoStop}%
\bibitem [{\citenamefont {Pan}\ \emph {et~al.}(2016)\citenamefont {Pan}, \citenamefont {Knox}, \citenamefont {Mulroe},\ and\ \citenamefont {Narimani}}]{Pan:2016zla}%
  \BibitemOpen
  \bibfield  {author} {\bibinfo {author} {\bibfnamefont {Z.}~\bibnamefont {Pan}}, \bibinfo {author} {\bibfnamefont {L.}~\bibnamefont {Knox}}, \bibinfo {author} {\bibfnamefont {B.}~\bibnamefont {Mulroe}}, \ and\ \bibinfo {author} {\bibfnamefont {A.}~\bibnamefont {Narimani}},\ }\href {\doibase 10.1093/mnras/stw833} {\bibfield  {journal} {\bibinfo  {journal} {Mon. Not. Roy. Astron. Soc.}\ }\textbf {\bibinfo {volume} {459}},\ \bibinfo {pages} {2513} (\bibinfo {year} {2016})},\ \Eprint {http://arxiv.org/abs/1603.03091} {arXiv:1603.03091 [astro-ph.CO]} \BibitemShut {NoStop}%
\bibitem [{\citenamefont {Green}\ and\ \citenamefont {Ridgway}(2020)}]{Green:2020fjb}%
  \BibitemOpen
  \bibfield  {author} {\bibinfo {author} {\bibfnamefont {D.}~\bibnamefont {Green}}\ and\ \bibinfo {author} {\bibfnamefont {A.}~\bibnamefont {Ridgway}},\ }\href {\doibase 10.1088/1475-7516/2020/12/050} {\bibfield  {journal} {\bibinfo  {journal} {JCAP}\ }\textbf {\bibinfo {volume} {12}},\ \bibinfo {pages} {050} (\bibinfo {year} {2020})},\ \Eprint {http://arxiv.org/abs/2008.05026} {arXiv:2008.05026 [astro-ph.CO]} \BibitemShut {NoStop}%
\bibitem [{\citenamefont {Blas}\ \emph {et~al.}(2011)\citenamefont {Blas}, \citenamefont {Lesgourgues},\ and\ \citenamefont {Tram}}]{Blas:2011rf}%
  \BibitemOpen
  \bibfield  {author} {\bibinfo {author} {\bibfnamefont {D.}~\bibnamefont {Blas}}, \bibinfo {author} {\bibfnamefont {J.}~\bibnamefont {Lesgourgues}}, \ and\ \bibinfo {author} {\bibfnamefont {T.}~\bibnamefont {Tram}},\ }\href {\doibase 10.1088/1475-7516/2011/07/034} {\bibfield  {journal} {\bibinfo  {journal} {JCAP}\ }\textbf {\bibinfo {volume} {07}},\ \bibinfo {pages} {034} (\bibinfo {year} {2011})},\ \Eprint {http://arxiv.org/abs/1104.2933} {arXiv:1104.2933 [astro-ph.CO]} \BibitemShut {NoStop}%
\bibitem [{\citenamefont {Venditti}\ \emph {et~al.}(2025)\citenamefont {Venditti}, \citenamefont {Munoz}, \citenamefont {Bromm}, \citenamefont {Fujimoto}, \citenamefont {Finkelstein},\ and\ \citenamefont {Chisholm}}]{Venditti:2025mgi}%
  \BibitemOpen
  \bibfield  {author} {\bibinfo {author} {\bibfnamefont {A.}~\bibnamefont {Venditti}}, \bibinfo {author} {\bibfnamefont {J.~B.}\ \bibnamefont {Munoz}}, \bibinfo {author} {\bibfnamefont {V.}~\bibnamefont {Bromm}}, \bibinfo {author} {\bibfnamefont {S.}~\bibnamefont {Fujimoto}}, \bibinfo {author} {\bibfnamefont {S.~L.}\ \bibnamefont {Finkelstein}}, \ and\ \bibinfo {author} {\bibfnamefont {J.}~\bibnamefont {Chisholm}},\ }\href@noop {} {\  (\bibinfo {year} {2025})},\ \Eprint {http://arxiv.org/abs/2505.20263} {arXiv:2505.20263 [astro-ph.GA]} \BibitemShut {NoStop}%
\bibitem [{\citenamefont {{DeBoer}}\ \emph {et~al.}(2017)\citenamefont {{DeBoer}}, \citenamefont {{Parsons}}, \citenamefont {{Aguirre}}, \citenamefont {{Alexander}}, \citenamefont {{Ali}} \emph {et~al.}}]{deboer17}%
  \BibitemOpen
  \bibfield  {author} {\bibinfo {author} {\bibfnamefont {D.~R.}\ \bibnamefont {{DeBoer}}}, \bibinfo {author} {\bibfnamefont {A.~R.}\ \bibnamefont {{Parsons}}}, \bibinfo {author} {\bibfnamefont {J.~E.}\ \bibnamefont {{Aguirre}}}, \bibinfo {author} {\bibfnamefont {P.}~\bibnamefont {{Alexander}}}, \bibinfo {author} {\bibfnamefont {Z.~S.}\ \bibnamefont {{Ali}}},  \emph {et~al.},\ }\href {\doibase 10.1088/1538-3873/129/974/045001} {\bibfield  {journal} {\bibinfo  {journal} {\pasp}\ }\textbf {\bibinfo {volume} {129}},\ \bibinfo {pages} {045001} (\bibinfo {year} {2017})},\ \Eprint {http://arxiv.org/abs/1606.07473} {arXiv:1606.07473 [astro-ph.IM]} \BibitemShut {NoStop}%
\bibitem [{\citenamefont {{HERA Collaboration}}(2023)}]{hera23}%
  \BibitemOpen
  \bibfield  {author} {\bibinfo {author} {\bibnamefont {{HERA Collaboration}}},\ }\href {\doibase 10.3847/1538-4357/acaf50} {\bibfield  {journal} {\bibinfo  {journal} {\apj}\ }\textbf {\bibinfo {volume} {945}},\ \bibinfo {eid} {124} (\bibinfo {year} {2023})},\ \Eprint {http://arxiv.org/abs/2210.04912} {arXiv:2210.04912 [astro-ph.CO]} \BibitemShut {NoStop}%
\bibitem [{\citenamefont {Bacon}\ \emph {et~al.}(2020)\citenamefont {Bacon} \emph {et~al.}}]{SKA:2018ckk}%
  \BibitemOpen
  \bibfield  {author} {\bibinfo {author} {\bibfnamefont {D.~J.}\ \bibnamefont {Bacon}} \emph {et~al.} (\bibinfo {collaboration} {SKA}),\ }\href {\doibase 10.1017/pasa.2019.51} {\bibfield  {journal} {\bibinfo  {journal} {Publ. Astron. Soc. Austral.}\ }\textbf {\bibinfo {volume} {37}},\ \bibinfo {pages} {e007} (\bibinfo {year} {2020})},\ \Eprint {http://arxiv.org/abs/1811.02743} {arXiv:1811.02743 [astro-ph.CO]} \BibitemShut {NoStop}%
\bibitem [{\citenamefont {{Braun}}\ \emph {et~al.}(2019)\citenamefont {{Braun}}, \citenamefont {{Bonaldi}}, \citenamefont {{Bourke}}, \citenamefont {{Keane}},\ and\ \citenamefont {{Wagg}}}]{SKA19}%
  \BibitemOpen
  \bibfield  {author} {\bibinfo {author} {\bibfnamefont {R.}~\bibnamefont {{Braun}}}, \bibinfo {author} {\bibfnamefont {A.}~\bibnamefont {{Bonaldi}}}, \bibinfo {author} {\bibfnamefont {T.}~\bibnamefont {{Bourke}}}, \bibinfo {author} {\bibfnamefont {E.}~\bibnamefont {{Keane}}}, \ and\ \bibinfo {author} {\bibfnamefont {J.}~\bibnamefont {{Wagg}}},\ }\href {\doibase 10.48550/arXiv.1912.12699} {\bibfield  {journal} {\bibinfo  {journal} {arXiv e-prints}\ ,\ \bibinfo {eid} {arXiv:1912.12699}} (\bibinfo {year} {2019})},\ \Eprint {http://arxiv.org/abs/1912.12699} {arXiv:1912.12699 [astro-ph.IM]} \BibitemShut {NoStop}%
\bibitem [{\citenamefont {{Liu}}\ and\ \citenamefont {{Shaw}}(2020)}]{liu20}%
  \BibitemOpen
  \bibfield  {author} {\bibinfo {author} {\bibfnamefont {A.}~\bibnamefont {{Liu}}}\ and\ \bibinfo {author} {\bibfnamefont {J.~R.}\ \bibnamefont {{Shaw}}},\ }\href {\doibase 10.1088/1538-3873/ab5bfd} {\bibfield  {journal} {\bibinfo  {journal} {\pasp}\ }\textbf {\bibinfo {volume} {132}},\ \bibinfo {eid} {062001} (\bibinfo {year} {2020})},\ \Eprint {http://arxiv.org/abs/1907.08211} {arXiv:1907.08211 [astro-ph.IM]} \BibitemShut {NoStop}%
\bibitem [{\citenamefont {Hunter}(2007)}]{Hunter:2007mat}%
  \BibitemOpen
  \bibfield  {author} {\bibinfo {author} {\bibfnamefont {J.}~\bibnamefont {Hunter}},\ }\href {\doibase 10.1109/MCSE.2007.55} {\bibfield  {journal} {\bibinfo  {journal} {Comput. Sci. Eng.}\ }\textbf {\bibinfo {volume} {9}},\ \bibinfo {pages} {90} (\bibinfo {year} {2007})}\BibitemShut {NoStop}%
\bibitem [{\citenamefont {Harris}\ \emph {et~al.}(2020)\citenamefont {Harris} \emph {et~al.}}]{Harris:2020xlr}%
  \BibitemOpen
  \bibfield  {author} {\bibinfo {author} {\bibfnamefont {C.}~\bibnamefont {Harris}} \emph {et~al.},\ }\href {\doibase 10.1038/s41586-020-2649-2} {\bibfield  {journal} {\bibinfo  {journal} {Nature}\ }\textbf {\bibinfo {volume} {585}},\ \bibinfo {pages} {3572} (\bibinfo {year} {2020})},\ \Eprint {http://arxiv.org/abs/2006.10256} {arXiv:2006.10256 [cs.MS]} \BibitemShut {NoStop}%
\bibitem [{\citenamefont {Virtanen}\ \emph {et~al.}(2020)\citenamefont {Virtanen} \emph {et~al.}}]{Virtanen:2019joe}%
  \BibitemOpen
  \bibfield  {author} {\bibinfo {author} {\bibfnamefont {P.}~\bibnamefont {Virtanen}} \emph {et~al.},\ }\href {\doibase 10.1038/s41592-019-0686-2} {\bibfield  {journal} {\bibinfo  {journal} {Nat. Methods}\ }\textbf {\bibinfo {volume} {17}},\ \bibinfo {pages} {261} (\bibinfo {year} {2020})},\ \Eprint {http://arxiv.org/abs/1907.10121} {arXiv:1907.10121 [cs.MS]} \BibitemShut {NoStop}%
\end{thebibliography}%
\end{document}